\documentclass{emulateapj}
\usepackage{graphicx}
\usepackage{epstopdf}
\begin{document}
\shorttitle{Retired A Stars and Their Companions. III.}
\title{Retired A Stars and Their Companions.  III.  \\ Comparing the Mass-Period Distributions of Planets Around \\ A-Type Stars and Sun-Like Stars$^1$}
\author{Brendan P.  Bowler,\altaffilmark{2} 
John Asher Johnson,\altaffilmark{2, 3} 
Geoffrey W. Marcy,\altaffilmark{4} 
Gregory W. Henry,\altaffilmark{5} 
Kathryn M. G. Peek,\altaffilmark{4} 
Debra A. Fischer,\altaffilmark{6} 
Kelsey I. Clubb,\altaffilmark{6} 
Michael C. Liu,\altaffilmark{2} 
Sabine Reffert,\altaffilmark{7} 
Christian Schwab,\altaffilmark{7} 
Thomas B. Lowe\altaffilmark{8}}

\begin{abstract}

We present an analysis of $\sim$5 years of Lick Observatory radial velocity measurements targeting a uniform sample of 31 intermediate-mass subgiants (1.5 $\lesssim$ $M_*/M_{\odot}$ $\lesssim$ 2.0) with the goal of measuring the occurrence rate of Jovian planets around (evolved) A-type stars and comparing the distributions of their orbital and physical characteristics to those of planets around Sun-like stars.  We provide updated orbital solutions incorporating new radial velocity measurements for five known planet-hosting stars in our sample; uncertainties in the fitted parameters are assessed using a Markov Chain Monte Carlo method.   The frequency of Jovian planets interior to 3 AU is 26$^{+9}_{-8}$\%, which is significantly higher than the ~5-10\% frequency observed around solar-mass stars.  The median detection threshold for our sample includes minimum masses down to \{0.2, 0.3, 0.5, 0.6, 1.3\} $M_\mathrm{Jup}$ within \{0.1, 0.3, 0.6, 1.0, 3.0\} AU.  To compare the properties of planets around intermediate-mass stars to those around solar-mass stars we synthesize a population of planets based on the parametric relationship $dN$ $\propto$ $M^{\alpha}P^{\beta}$$d$ln$M$$d$ln$P$, the observed planet frequency, and the detection limits we derived.  We find that the values of $\alpha$ and $\beta$ for planets around solar-type stars from Cumming et al. fail to reproduce the observed properties of planets in our sample at the 4 $\sigma$ level, even when accounting for the different planet occurrence rates.  Thus, the properties of planets around A stars are markedly different than those around Sun-like stars, suggesting that only a small ($\sim$ 50\%) increase in stellar mass has a large influence on the formation and orbital evolution of planets.

\end{abstract}

\altaffiltext{1}{Based on observations obtained at the Lick Observatory, which is operated by the University of California}

\altaffiltext{2}{Institute for Astronomy, University of Hawai`i, 2680 Woodlawn Drive, Honolulu, HI 96822, USA; \texttt{bpbowler@ifa.hawaii.edu}}
\altaffiltext{3}{Current address: Department of Astrophysics, California Institute of Technology, MC 249-17, Pasadena, CA 91125, USA}
\altaffiltext{4}{Department of Astronomy, MS 3411, University of California, Berkeley, CA 94720-3411, USA}
\altaffiltext{5}{Center of Excellence in Information Systems, Tennessee State University, 3500 John A. Merritt Blvd., Box 9501, Nashville, TN 37209, USA}
\altaffiltext{6}{Department of Physics and Astronomy, San Francisco State University, San Francisco, CA 94132, USA}
\altaffiltext{7}{ZAH-Landessternwarte, K\"{o}nigstuhl 12, 69117 Heidelberg, Germany}
\altaffiltext{8}{UCO/Lick Observatory, Santa Cruz, CA 95064}

\section{Introduction}

Our understanding of planet formation has rapidly improved over the past 15 years.  Prior to the discovery of the first extrasolar planet orbiting a solar-type star (51 Peg b; \citealt{Mayor:1995p18405}), it was widely assumed that extrasolar giant planet semimajor axes would mimic those of the gas giants in our own solar system, which orbit at distances $>$ 5 AU.  In the years that followed it became apparent that an \emph{in situ} formation model was not universally applicable because radial velocity surveys were finding Jovian planets in abundance well inside the canonical ice line.

\begin{figure}
%     \plotone{f1.eps}
       \resizebox{3.5in}{!}{\includegraphics{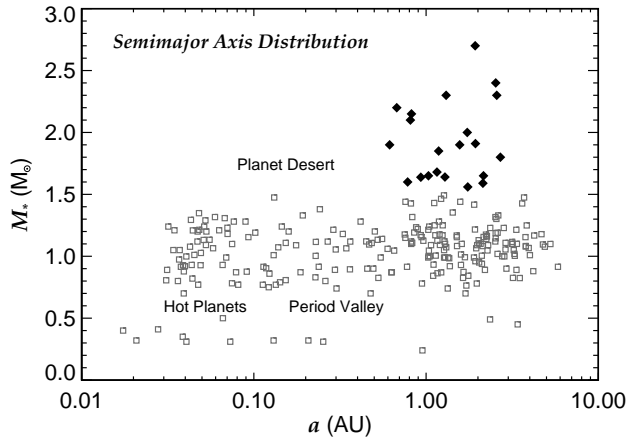}}
  \caption{The semimajor axis distribution of exoplanets within 200 pc.  The data are compiled from the literature and are maintained by the California \& Carnegie Planet Search team.  Planet-hosting stars with masses less than 1.5 $M_{\odot}$ are shown as open squares, while planet-hosting stars with higher masses are shown as filled diamonds. For solar-type stars, planets can be grouped into two populations: ``hot planets'' located at distances $\lesssim$ 0.1 AU and longer period planets located at distance $\gtrsim$ 1 AU.  Between these two groups lies a relative dearth of planets in a region known as the ``period valley.''  No planets have been discovered with semimajor axes $<$ 0.6 AU for stellar masses $>$ 1.5 $M_\mathrm{\sun}$, creating a ``planet desert'' in that region.  Planets with semimajor axes $\gtrsim$ 5 AU are limited by the current baseline of radial velocity observations.     \label{f1.eps} } 
\end{figure}

Over 350 planets have now been discovered, 282 of which reside around stars within 200 pc.\footnote{As of August 2009; see http://exoplanets.org.} Sufficiently large samples are available for the statistical properties of exoplanets to reveal themselves, providing information about  the planet formation and migration processes.   For solar-type stars (F, G, and K dwarfs), Jovian planets fall into two rough populations:  ``hot planets'' with $a$ $\lesssim$ 0.1 AU (\citealt{Fischer:2008p10236}) and those that orbit beyond $\sim$ 1 AU (Figure \ref{f1.eps}).  These observations were explained \emph{a posteriori} in terms of orbital migration (\citealt{Papaloizou:2007p8045}) and planet-planet scattering (\citealt{Nagasawa:2008p18396}; \citealt{Marchi:2009p17699}; \citealt{Ford:2008p19033}), with the dearth of planets with periods between $\sim$ 10-100 days (the ``period valley'') possibly resulting from differential mass-dependent orbital migration (\citealt{Udry:2003p18397}; \citealt{Burkert:2007p10310}; \citealt{Currie:2009p15637}).

While much is known about planets around Sun-like stars (e.g., \citealt{Butler:2006p3743}; \citealt{Udry:2007p18401}; \citealt{Marcy:2008p10230}; \citealt{Johnson:2009p18532}; \citealt{Wright:2009p18840}), comparatively little is known about planets around intermediate-mass (IM) stars with $M_*$ $>$ 1.5 $M_{\odot}$.  Main sequence A- and F-type stars are problematic Doppler targets because of their high jitter levels and rotationally-broadened absorption features (\citealt{Galland:2005p18305}).  On the other hand, evolved IM stars have lower jitter levels as well as narrower and more numerous absorption lines resulting from their slow rotation and cool photospheres.  As a consequence, nearly all radial velocity surveys of IM stars are targeting evolved G- or K-type subgiants and giants (\citealt{Frink:2002p9370}, \citealt{Setiawan:2003p9372}, \citealt{Sato:2005p18303}; \citealt{Johnson:2006p170}, \citealt{Niedzielski:2007p17717}, \citealt{Dollinger:2007p18020}, \citealt{Lovis:2007p17712}, \citealt{Liu:2009p17779}), although at least one survey is targeting their main sequence progenitors (\citealt{Galland:2005p18305}; \citealt{Lagrange:2009p15653}).

Recently \citet{Johnson:2007p165} showed that planets orbiting evolved A-type stars have large semimajor axes compared to planets around solar-type stars.  This trend has become even stronger with the discovery of more planets orbiting IM stars (\citealt{Sato:2008p9941}).  Specifically, over 20 planets have been discovered around stars with minimum masses $>$ 1.5 $M_{\sun}$ but none have semimajor axes $<$ 0.6 AU.  This ``planet desert'' is shown in Figure \ref{f1.eps} and a summary of known planets around IM stars is presented Table \ref{evolvedtab}.

\begin{deluxetable*}{lcccccccc}
%\rotate
\tabletypesize{\scriptsize}
\tablewidth{0pt}
\tablecolumns{8}
\tablecaption{Planetary-Mass Companions to Evolved Intermediate-Mass Stars with $M_*$ $>$ 1.5 $M_{\odot}$ \label{evolvedtab}}
\tablehead{
        \colhead{}   &    \colhead{$M_*$}   & \colhead{}    & \colhead{$R_*$}    & \colhead{$M_\mathrm{P}$ sin$i$}    & \colhead{$a$}  &  \colhead{}   &  \colhead{}  \\
        \colhead{Star}   &    \colhead{($M_{\odot}$)}   & \colhead{SpT}    & \colhead{($R_{\odot}$)}    & \colhead{($M_\mathrm{Jup}$)}    & \colhead{(AU)}  &  \colhead{$e$}  & \colhead{Ref} 
}
\startdata
HD 13189               &          2-6               &   K2 II        &  \nodata                            &  8-20                    &  1.5-2.2            &   0.27 (0.06)         &  1, 2 \\
$\epsilon$ Tau       &    2.7 (0.1)            &  K0 III        &   13.7 (0.6)                       &     7.6 (0.2)          &   1.93 (0.03)    &   0.151 (0.023)     &   3  \\
NGC 2423 No. 3    &    2.4 (0.2)            &   \nodata   &   \nodata                          &      10.6                &   2.10                &    0.21 (0.07)        &   4   \\
81 Cet                      &     2.4 (2.0-2.5)    &  G5 III:       &    11 (10-13)                    &      5.3                  &    2.5                 &    0.206 (0.029)    &   5   \\
HD 104985             &   2.3                      &  G9 III        &   11                                   &   8.3                      &  0.95                 &   0.090 (0.009)    & 6, 7 \\
18 Del                      &     2.3                    &   G6 III       &   8.5                                  &      10.3                &     2.6                 &    0.08 (0.01)        &   6   \\
HD 17092               &   2.3 (0.3)             &  K0 III         &   10.9 (2.8)                      &    4.6 (0.3)           &   1.29 (0.05)    &     0.166 (0.052)   &   8 \\
$\xi$ Aql                  &    2.2                     &    K0 III        &   12                                  &        2.8                 &     0.68             &   0.0 (fixed)           &   6   \\
14 And                     &      2.2 (2.0-2.3)   &  K0 III         &    11 (10-12)                   &    4.8                     &  0.83                &    0.0 (fixed)          &   5   \\
HD 81688               &    2.1                     &    K0III-IV    &   13                                  &       2.7                  &     0.81             &   0.0 (fixed)           &   6   \\
HD 173416             &    2.0 (0.3)           & G8 III          &    13.5 (0.9)                      &    2.7 (0.3)           &   1.16 (0.06)    &    0.21 (0.04)       &   9  \\
HD 11977               &  1.91 (0.21)         &    G5 III       &   10.09 (0.32)                  &   6.54                   &   1.93               &   0.40 (0.07)        &  10, 11 \\
HD 102272             &      1.9 (0.3)         &  K0 III         &   10.1 (4.6)                       &   5.9 (0.2)            & 0.614 (0.001) &  0.05 (0.04)         &  12   \\
\ \ \ \ \ \ \ \ \ \      ''        &            ''                 &         ''         &              ''                            &    2.6 (0.4)           &  1.57 (0.05)     &    0.68 (0.06)       &  12   \\
$\beta$ Gem          &  1.86, 1.7 (0.4)    &   K0 III        &     8.8 (0.1)                       &   2.9 (0.3)            &   1.69 (0.03)    &    0.06 (0.04)       &     13, 14, 15, 16    \\
HD 89744               &    1.86 (0.18)       &  F7 IV         &    2.08 (0.06)                   &     7.2                   &    0.88               &     0.70 (0.02)      &   17, 18 \\
HD 210702             &     1.85 (0.13)      &   K1 IV       &    4.45 (0.07)                  & 1.97 (0.11,0.18) & 1.20 (0.02,0.03)  & 0.036 ($<$0.106) &  19, 20  \\
$\kappa$ CrB         &      1.84 (0.13)     &   K0 IV       &   4.71 (0.08)                   & 2.01 (0.11,0.17) & 2.80 (0.07,0.08) & 0.044 ($<$0.123) &  21, 20      \\
6 Lyn                        &     1.82 (0.13)   &   K0 IV       &    5.2 (4.9-5.6)                 &  2.21 (0.11,0.16) & 2.18 (0.05,0.06) &  0.059 ($<$0.125) &   20, 5   \\
HD 167042             &      1.72 (0.12)     &    K1 IV     &   4.30 (0.07)                     &  1.70 (0.09,0.12) & 1.32 (0.03,0.04) & 0.089 (0.028,0.065) &  21, 20, 5  \\
HD 192699             &      1.69 (0.12)     &   G8 IV      &    3.90 (0.06)                   &  2.40 (0.15,0.21) & 1.15 (0.02,0.03) &  0.129 (0.029,0.060) &  19, 20  \\
HD 175541             &      1.65 (0.12)    &   G8 IV       &    3.80 (0.09)                   &  0.70 (0.06,0.08) & 1.03 (0.02,0.03) & 0.083 ($<$0.283) &  19   \\
HD 5319                  &       1.59 (0.18)   &    G5 IV      &    3.26 (0.50,0.41)         &   1.94                     &   1.75                   &  0.12 (0.08)           &  22

\enddata
\tablerefs{(1) \citet{Hatzes:2005p10039};  (2) \citet{Schuler:2005p17780};  (3) \citet{Sato:2007p9819};  (4) \citet{Lovis:2007p17712};  (5) \citet{Sato:2008p14857};  (6)  \citet{Sato:2008p9941};  (7) \citet{Sato:2003p9498};  (8) \citet{Niedzielski:2007p17717};  (9) \citet{Liu:2009p17779};  (10) \citet{Setiawan:2005p9575};  (11) \citet{daSilva:2006p10898};  (12) \citet{Niedzielski:2009p15626};  (13) \citet{Hatzes:2006p10066};  (14) \citet{Nordgren:2001p17715};  (15) \citet{AllendePrieto:1999p17714};  (16) \citet{Reffert:2006p10079};  (17) \citet{Korzennik:2000p17781};  (18) \citet{Valenti:2005p11833};  (19) \citet{Johnson:2007p165};  (20) this work;  (21) \citet{Johnson:2008p166}; (22) \citet{Robinson:2007p18408}}
\tablecomments{HD 47536 b is omitted because the most likely stellar mass is between 1.0-1.5 $M_{\sun}$ (\citealt{Setiawan:2003p9372}).  $\gamma$ Cep A b (\citealt{Hatzes:2003p10013}) is omitted because the stellar mass estimate was revised to 1.4 $M_{\sun}$ (\citealt{Neuhauser:2007p18407}).   The minimum mass of the companion to 11 Comae is 19.4 $M_\mathrm{Jup}$ (\citealt{Liu:2008p17989}) and is therefore probably a brown dwarf.  References are for planet discoveries and host-star physical properties.  Additionally, HD 90043 and HD 200964 are intermediate-mass planet-hosting stars, but do not yet have unique orbit solutions and so were omitted from this table.}

\end{deluxetable*}

There are several explanations for this observational result.  One possibility is that the swollen radii of evolved stars have engulfed or tidally disrupted short-period planets.  The current census of IM exoplanet-host stars mostly consists of red clump giants (core helium and hydrogen shell burning post-RGB stars) and subgiants (stars with contracting inert helium cores and hydrogen-burning shells).   Planet-hosting stars in the clump giant phase have radii between $\sim$ 8-14 $R_{\sun}$ ($\sim$ 0.03-0.06 AU) so any planets orbiting inside $\sim$ 0.1 AU will likely have been engulfed by an expanding radius during post-main sequence stellar evolution.  However, a tidal torque from an expanding stellar surface can also decay the orbits of short period planets.   \citet{Sato:2008p9941} numerically trace the semimajor axis evolution of short-period planets around evolving RGB stars and show that, in their past, IM clump giants may have engulfed or disrupted the orbits of planets out to $\sim$ 0.5 AU.   With radii between $\sim$ 2-5 $R_{\sun}$ ($\sim$ 0.01-0.02 AU), planets orbiting IM subgiants are the least affected by stellar evolution.  Stellar evolution may therefore explain the lack of short-period planets around IM clump giants, but the same result for IM subgiants suggests that the observed trend is not due to post-main sequence engulfment.

The observed semimajor axis distribution can also be explained as a result of inward orbital migration combined with mass-dependent disk dispersal lifetime.  The rocky progenitors of Jovian planets ($\sim$10 $M_{\earth}$ cores) can form at distances $\gtrsim$ 8 AU for stellar masses between $\sim$ 1.5-3 $M_{\sun}$ (\citealt{Kennedy:2008p18349}).  \citet{Currie:2009p15637} performed Monte Carlo simulations of Jovian planet formation and migration around IM stars and, using simple stellar mass-dependent gas disk lifetime relations (\citealt{Kennedy:2009p18354}) and Type II migration models (\citealt{Ida:2004p18000}), was able to successfully reproduce the observed dearth of short-period planets.  In this scenario inward migration is halted once rapid disk dispersal occurs, stranding migrating planets at semimajor axes that depend on stellar mass.  

An alternative explanation was offered by \citet{Kretke:2008p14791}.  They found that the protoplanetary disks of young IM stars will develop a maximum surface density at $\sim$ 1 AU as a result of magnetrotational instability of the inner disk, leading to a trapping and accumulation of solids that can then grow to form rocky cores and Jovian planets.  This formation scenario provides a mechanism for \emph{in situ} formation of Jovian planets interior to the ice line, which is located near 3 AU at 10 Myr for a 2 $M_{\sun}$ star (\citealt{Kennedy:2008p18349}).

Models of Jovian planet formation around IM stars make few quantitative predictions that can be observationally tested.  Simple disk depletion plus migration models for IM planet hosts (1.5-3.0 $M_{\sun}$) by \citet{Currie:2009p15637} predict occurrence rates for Jovian planets with semimajor axes $<$ 0.5 AU to be $\lesssim$ 1.5\% and for Jovian planets with semimajor axes $>$ 0.5 AU to be $\gtrsim$ 7.5\%.  \citet{Kennedy:2008p18349} use a semianalytic model of protoplanetary disk evolution to study snow line locations and planet formation rates around stars of varying masses.  For stellar masses between 1.5-2.0 $M_{\sun}$, their models predict that Jovian planet occurrence rates reach frequencies of $\sim$10-15\%.   \citet{Kretke:2008p14791} suggest that multiple planetary systems may form more efficeintly around IM stars compared to other stellar mass regimes.

The semimajor axis distribution of planets with minimum masses between $\sim$ 2-10 $M_\mathrm{Jup}$ orbiting IM stars is beginning to be better constrained by observations, but little is known about planets with masses $<$ 1.5 $M_\mathrm{Jup}$.  This poor understanding is a direct result of the dearth of low-mass planets currently known, with only one having a minimum mass below 1.5 $M_\mathrm{Jup}$ (HD 175541b with $M_\mathrm{P}$sin$i$ = 0.70 $M_\mathrm{Jup}$).  It is unclear, however, whether this scarcity is a result of a detection bias caused by higher jitter levels in IM stars or whether it reflects an intrinsic shortage of low-mass planets.  Unfortunately, models of planet formation in this stellar mass regime have made few predictions about low-mass planetary companions.  For solar-type stars, \citet{Mayor:2009p17784} estimate the frequency of Neptune-mass planets with periods $<$ 50 days to be at least 30\%.  This raises the exciting possibility that low-mass planets could be abundant around IM stars, especially in light of recent studies suggesting that the frequency of Jovian-mass planets scales with stellar mass (\citealt{Johnson:2007p169}).  Testing these theories requires an understanding of the detection limits of radial velocity surveys.

The goal of this study is twofold: to derive the occurrence rate of Jovian planets around IM stars and to characterize the distributions of planet periods ($P$) and minimum masses ($M_\mathrm{P}$sin$i$).  In addition we take the opportunity to update the orbit solutions for known planet-hosting stars in our sample using new radial velocity measurements.  To address the aforementioned questions we make use of a uniform sample of 31 IM subgiants taken from an ongoing radial velocity survey at Lick Observatory (\citealt{Johnson:2006p170}; \citealt{Peek:2009p18409}).  Our observations span $\sim$ 5 years and sample semimajor axes out to several AU, enabling detailed comparisons with previous surveys targeting solar-type stars. 

In $\S$2 we describe our radial velocity measurements and define our sample selection.  Updated Keplerian orbits for five previously-known planetary systems are presented in $\S$3 and in $\S$4 we describe time-series photometric observations of these systems. In $\S$5 we derive detection limits for stars in our sample.  We discuss the frequency of Jovian-mass planets in $\S$6, and  in $\S$7 we compare the mass-period distribution for solar-type stars to the results from our sample of IM subgiants.  Finally, we discuss the implications of our work in $\S$8.

\section{Radial Velocity Observations}

\begin{figure}
  \resizebox{3.5in}{!}{\includegraphics{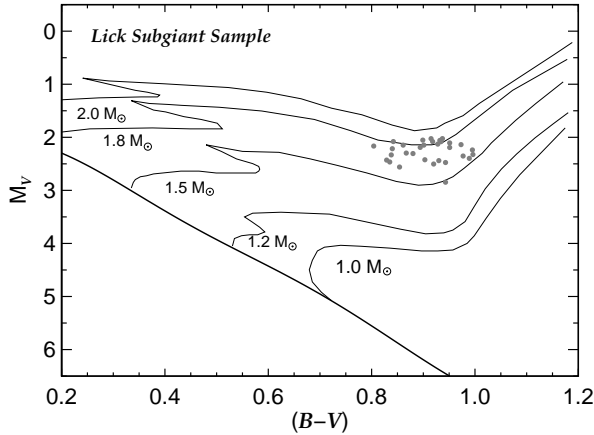}}
    %\plotone{f2.eps}
  \caption{Color-magnitude diagram showing the location of our subsample of subgiants monitored at Lick Observatory.  The thick line shows the locus of $Hipparcos$ main sequence stars and the thin lines display the solar metallicity evolutionary tracks of \citet{Girardi:2002p18593}.  Our uniform sample of 31 subgiants is a subset of the larger Lick Subgiant Planet Search program.  \label{f2.eps} } 
\end{figure}

Our sample of subgiants is derived from a larger planet-hunting program at Lick Observatory targeting evolved stars (the Lick Subgiant Planet Search: \citealt{Johnson:2006p170}; \citealt{Peek:2009p18409}).  The original survey consists of 159 stars selected on the basis of $Hipparcos$ $B-V$ colors and $M_V$ absolute magnitudes such that they lie one magnitude above the main sequence, have masses $\gtrsim$ 1.2 $M_{\sun}$, and avoid red giant branch (RGB), clump giant, and Cepheid variable regions of the HR diagram.  Known spectroscopic and proper motion binaries within $\sim$ 2$\arcsec$ are excluded. To obtain a more uniform population for this study we selected a sample of stars from the original survey based on the following criteria: 2 $<$ $M_V$ $<$ 3, 0.8 $<$ $B$ -- $V$ $<$ 1.0, $M_*$ $>$ 1.5 $M_{\sun}$, and at least four radial velocity measurements (the average number of measurements in our final sample is 26).\footnote{Only one star, HD 33066, has $<$ 4 radial velocity measurements, but it is a previously-known spectroscopic binary.}   The magnitude cuts exclude clump giants and solar-mass subgiants, while the color cuts exclude RGB stars and yellow stragglers, the latter of which have high jitter levels and large binary frequencies.  Altogether our sample includes 31 intermediate-mass subgiants (Figure \ref{f2.eps}).

Radial velocity measurements were obtained at Lick Observatory's 3 m Shane telescope and 0.6 m Coude Auxiliary Telescope with the Hamilton spectrometer (\citealt{Vogt:1987p18399}; $R$ $\sim$ 50,000 at $\lambda$ = 5500 \AA).  Doppler shifts are measured from each spectrum using the iodine cell method described in detail by \citet{Butler:1996p12444} and summarized as follows.  A temperature-controlled Pyrex cell containing gaseous iodine is placed at the entrance slit of the spectrometer.  The dense set of narrow molecular lines imprinted on each stellar spectrum from 5000 to 6000 \AA \ provides a robust wavelength scale for each observation, as well as information about the shape of the spectrometer's instrumental response (\citealt{Marcy:1992p18836}).  Additional details about the observations can be found in \citealt{Johnson:2007p165}.

\begin{figure}
%    \plotone{f3.eps}
           \resizebox{3.5in}{!}{\includegraphics{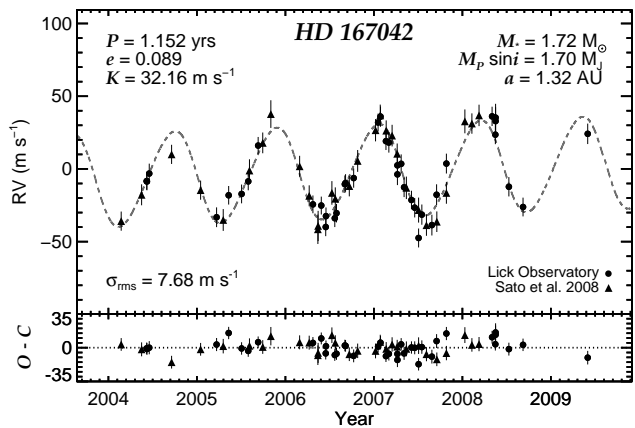}}
  \caption{Updated Keplerian orbit for HD 167042.  The gray dashed line shows the best-fitting orbital solution.  The lower panel displays the residuals after subtracting off the model.  There is evidence for an additional outer companion based on a linear trend of 2.14$^{+0.99}_{-0.73}$ m s$^{-1}$ yr$^{-1}$.  \label{f3.eps} } 
\end{figure}

\begin{figure}
%  \plotone{f4.eps}
             \resizebox{3.5in}{!}{\includegraphics{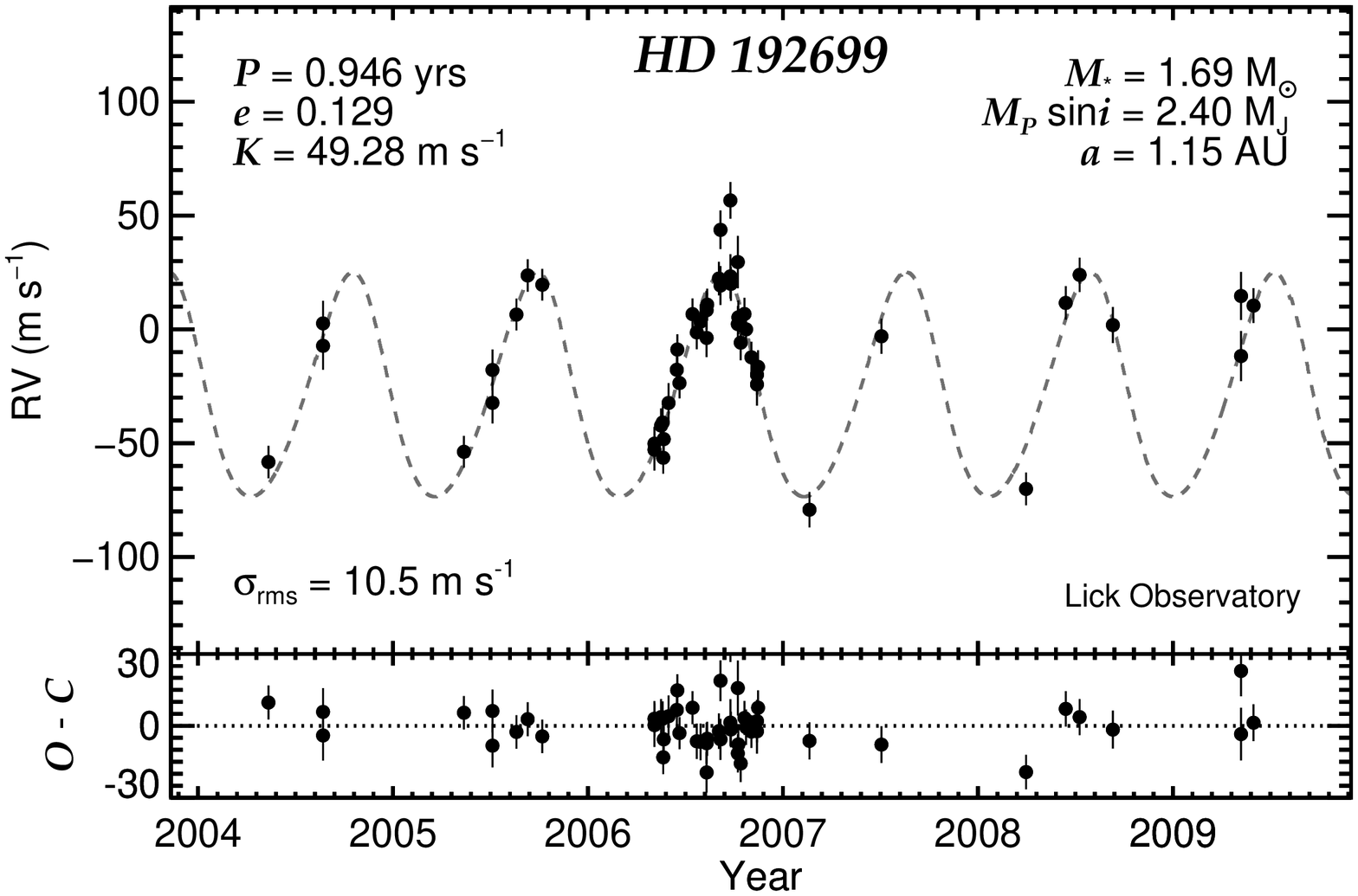}}
  \caption{Updated Keplerian orbit for HD 192699.  The gray dashed line shows the best-fitting orbital solution.  The lower panel displays the residuals after subtracting off the model.     \label{f4.eps} } 
\end{figure}

\begin{figure}
%  \plotone{f5.eps}
             \resizebox{3.5in}{!}{\includegraphics{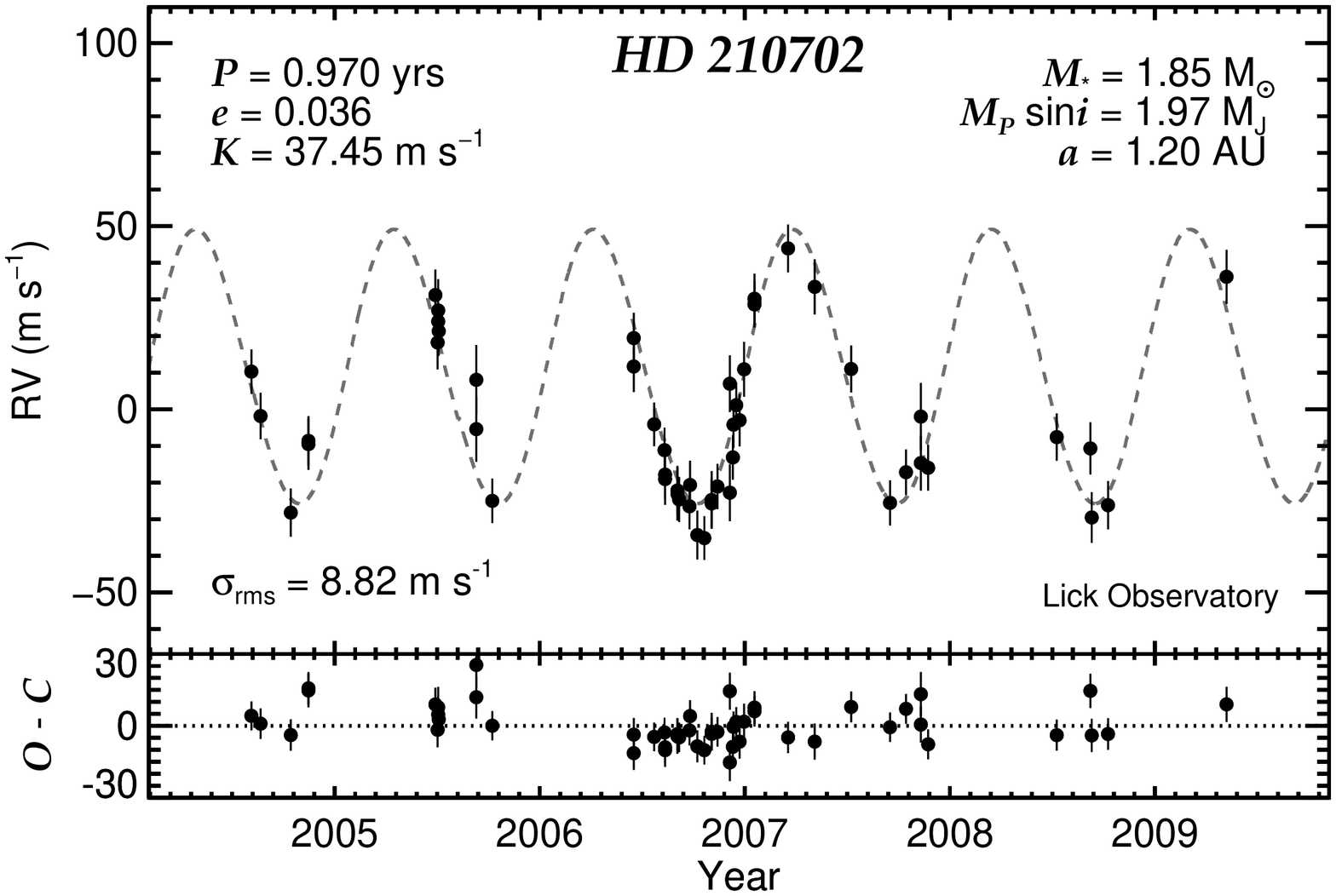}}
  \caption{Updated Keplerian orbit for HD 210702.  The gray dashed line shows the best-fitting orbital solution.  The lower panel displays the residuals after subtracting off the model.    \label{f5.eps} } 
\end{figure}

\begin{figure}
  %\plotone{f6.eps}
           \resizebox{3.5in}{!}{\includegraphics{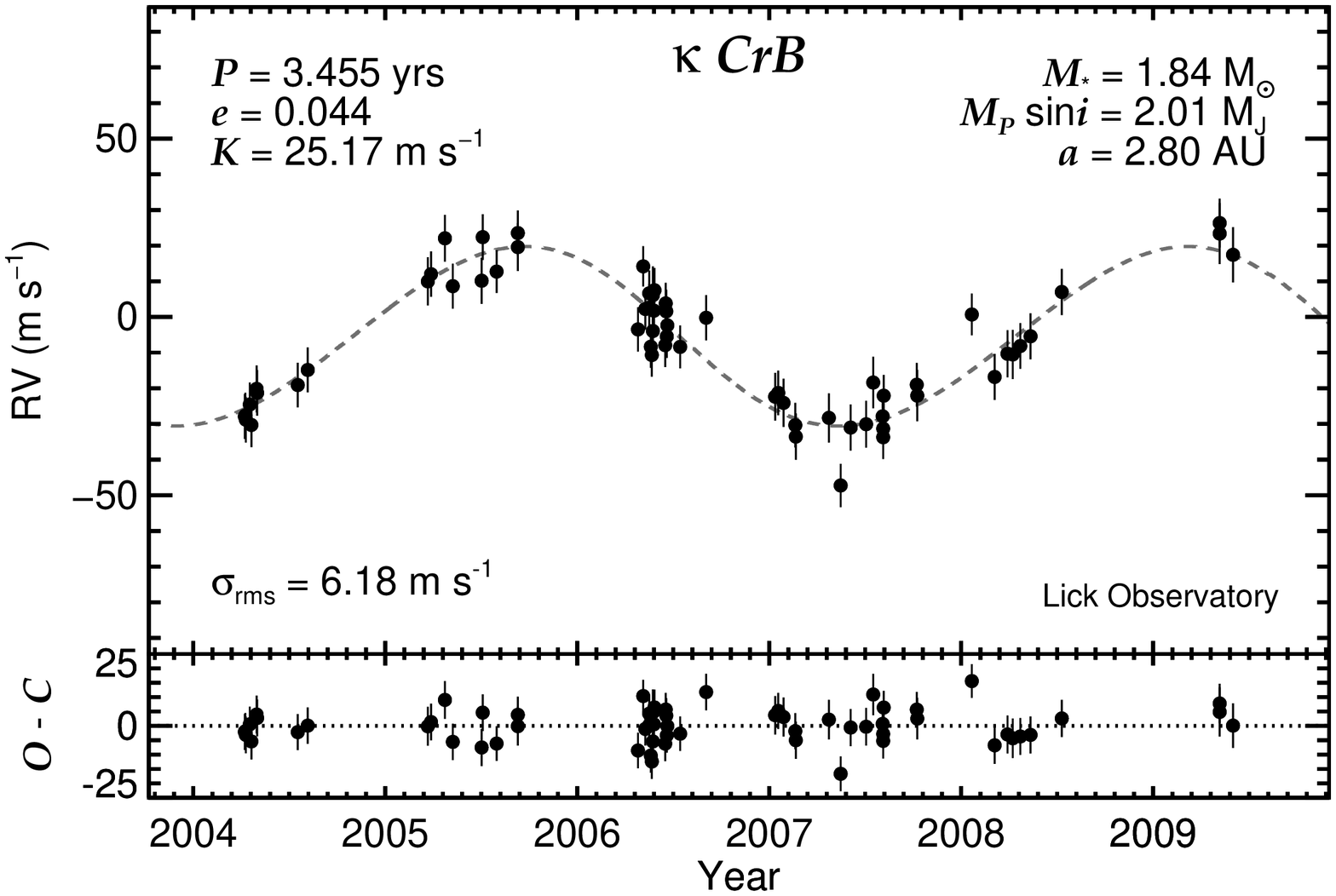}}
  \caption{Updated Keplerian orbit for $\kappa$ CrB (HD 142091).  The gray dashed line shows the best-fitting orbital solution.  The lower displays the residuals after subtracting off the model.     \label{f6.eps} } 
\end{figure}

\begin{figure}
%  \plotone{f7.eps}
           \resizebox{3.5in}{!}{\includegraphics{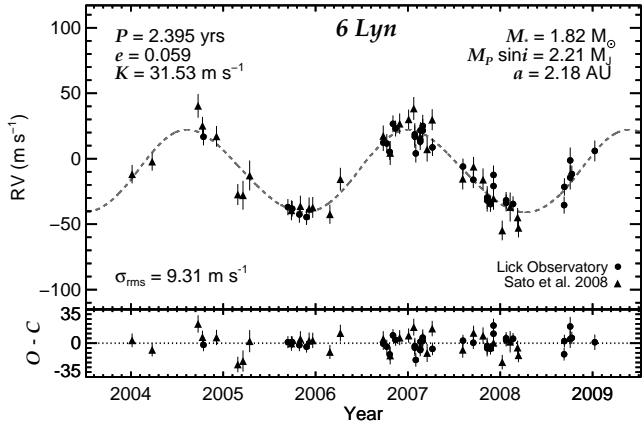}}
  \caption{Updated Keplerian orbit for 6 Lyn (HD 45410).  The gray dashed line shows the best-fitting orbital solution.  The lower panel displays the residuals after subtracting off the model.    \label{f7.eps} } 
\end{figure}

\section{Keplerian Orbits and Uncertainties}

We derive Keplerian orbital solutions for the five previously-known planets in our sample using \texttt{RVLIN}, an efficient orbit-fitting routine written in IDL and based on a partial linearization of Kepler's equations (\citealt{Wright:2009p18376}).  The radial velocity measurements were first combined into two-hour bins and the total measurement errors were estimated by adding 5 m s$^{-1}$ to the internal measurement errors, in quadrature, to account for the moderate jitter level of typical subgiants (\citealt{Fischer:2003p18410}; \citealt{Wright:2005p12980}; \citealt{Johnson:2007p165}).   We combine our observations of HD 167042 and 6 Lyn (HD 45410) with those made available by \citet{Sato:2008p14857} to create a larger data set for improved orbital phase coverage.  The data sets were merged by simultaneously fitting for the constant offset between the two observatories.

Long-period companions will manifest as constant accelerations in the radial velocity measurements.  To first order this drift can be modeled as a linear velocity trend for very long-period companions for which the period is much greater than the observing time baseline.  When the slope has an amplitude comparable to the noise it is often difficult to determine whether a trend should be included in an orbit solution.  The goodness of fit statistic (typically the $\chi^2$ value) might be reduced with the orbit-plus-trend solution, but, by that criterion for model selection, a more complicated model with multiple planets could be fit to a radial velocity set for which there is clearly insufficient evidence for additional companions. More generally, how does one decide whether including more parameters in a model is justified?  Fortunately this problem has received a great deal of attention in statistics and is the subject of Bayesian Model Comparison.\footnote{A popular frequentist approach to model selection is to derive false alarm probabilities for a more complicated model using of the $F$-test (see, e.g., \citealt{Cumming:1999p9255}).  We choose to follow a Bayesian approach in this work.}   A commonly used model selection tool is the Bayesian Information Criterion (BIC; \citealt{Schwarz:1978p17783}; \citealt{Liddle:2004p15221}; \citealt{Liddle:2007p18400}), which rewards better-fitting models but penalizes models that are overly complex:

\begin{equation}
\mathrm{BIC} \equiv -2 \ln \mathcal{L_\mathrm{max}} + k \ln N,
\end{equation}

\noindent where $\mathcal{L_\mathrm{max}}$ is the maximum likelihood for a particular model with $k$ free parameters and
$N$ data points.  In general the model with the smaller BIC value is preferred.

We use BIC values for orbit solutions with and without a velocity trend to decide whether to include a linear slope in the model.  The only planet-hosting star for which a trend produced a lower BIC value was HD 167042.  We take this as evidence for a long-period companion in that system.  The best-fitting orbits for the five known planet-hosting subgiants are shown in Figures \ref{f3.eps}-\ref{f7.eps}, and the updated orbital parameters are listed in Table \ref{orbitpars1}.

\begin{deluxetable*}{lccccc}
%\rotate
\tabletypesize{\scriptsize}
\tablewidth{0pt}
\tablecolumns{6}
\tablecaption{Updated Orbital Parameters \label{orbitpars1}}
\tablehead{
        \colhead{Parameter}   &    \colhead{HD 167042b}   &  \colhead{HD 192699b} &  \colhead{HD 210702b}  &   \colhead{$\kappa$ CrB b}  & \colhead{6 Lyn b}  
}
\startdata
$P$ (days)                                                          &  420.77 (3.48, 3.11)     &  345.53 (1.77, 1.63)         &  354.29 (2.31, 2.13)          &  1261.94 (28.91, 23.97)        & 874.774 (16.27, 8.47)               \\
$T_{\mathrm{p}}$ (JD -- 2,450,000)              &  4230.1 (40.6, 41.6)     &   4036.6 (21.0, 22.1)        &  4142.6 (78.3, 100.1)         &  3909.2 (332.5,  260.9)        &  4024.5 (180.2, 130.9)              \\
$e$                                                                      &  0.089 (0.028, 0.065)   &  0.129 (0.029, 0.060)         &  0.036 ($<$0.106)              &   0.044 ($<$0.123)             &  0.059 ($<$0.125)                   \\
$\omega$  ($^{\circ}$)                                     &  85.7 (35.4, 35.8)          &  41.3 (23.3, 25.9)             &  282.3 (82.3, 98.0)           &  148.4 (83.2, 94.2)         &  314.9 (75.5, 52.6)                    \\
$K$ (m s$^{-1}$)                                              &  32.16 (1.32, 1.32)        & 49.3 (2.6, 3.2)                   &  37.45 (1.90, 2.49)           &   25.17 (1.12, 1.55)       &  31.53 (1.12, 1.32)                     \\
$M_P$ sin$i$ ($M_\mathrm{Jup}$)             & 1.70 (0.09, 0.12)          & 2.40 (0.15, 0.21)               &  1.97 (0.11, 0.18)              & 2.01 (0.11, 0.17)          & 2.21 (0.11, 0.16)                       \\
$a$ (AU)                                                           &  1.32 (0.03, 0.04)         &  1.15 (0.02, 0.03)              &   1.20 (0.02, 0.03)             & 2.80 (0.07, 0.08)          & 2.18 (0.05, 0.06)                       \\
$\chi^2_{\nu}$                                                  &  1.17                              &  1.82                              &  1.49                                  & 1.01              &  1.43                                           \\
BIC (trend)                                                        &  \bf{99.84}                     & 104.97                            &   90.19                                &  82.51               &  110.25                                      \\
BIC (no trend)                                                  &  101.88                        &  \bf{101.72}                      &  \bf{89.18}                       & \bf{80.15}                  &  \bf{106.31}                          \\
$dv$/$dt$ (m s$^{-1}$ yr$^{-1}$)                 &  2.14 (0.99, 0.73)         &  0.0 (fixed)                       & 0.0 (fixed)                        &  0.0 (fixed)                & 0.0 (fixed)                            \\
\enddata
\tablecomments{Updated orbital parameters for planets orbiting intermediate-mass stars in our sample.  In parentheses the upper and lower boundaries of the regions encompassing 68.3\% of the posterior probability density distributions about the median are listed.  For low eccentricities, the peak of the probability density distribution may not be encompassed by the 68.3\% region about the median.  In those cases an upper limit encompassing 95.4\% of the data is quoted.}
\end{deluxetable*}

Minimum masses and semimajor axes are calculated using the analytic approximation for $M_\mathrm{P}$$\sin i$ as a function of the observed orbital parameters and Kepler's third law (i.e.,  Equations 2 and 3 in \citealt{Cumming:1999p9255}).  We use newly determined stellar masses following the method described in \citet{Johnson:2007p165} based on new [Fe/H] values (see Table \ref{evolvedtab}), which are derived from updated iodine-free template spectra following the iterative scheme from Figure 1 of \citet{Valenti:2009p18837}.

We use a Markov-Chain Monte Carlo (MCMC) algorithm following the description in \citet{Ford:2005p12728} to estimate our
orbital parameter uncertainties (see also \citealt{Ford:2006p14388}, \citealt{Ford:2007p18299}).  MCMC is a Bayesian inference technique that
uses the data to explore the shape of the likelihood function for each parameter
of an input model.  This method of error analysis has been shown to be more efficient and accurate than bootstrap Monte Carlo resampling techniques (\citealt{Ford:2005p12728}).  Every step in the chain represents a variation of the period
($P$), eccentricity, ($e$), time of periastron passage ($T_\mathrm{P}$),  argument
of periastron ($\omega$), velocity semi-amplitude ($K$),  and constant velocity offset.  A velocity trend parameter (d$v$/dt) is allowed to vary for HD 167042.  Additionally, for HD 167042 and 6 Lyn, a constant radial velocity offset between Lick Observatory measurements and those from \citet{Sato:2008p14857} is treated as a free parameter to account for the systematic offset between observations from different telescopes.
We allow one random parameter to be altered at each step by drawing a new value from a
Gaussian transition distribution.\footnote{Similar to, but not to be confused with, Gibbs sampling (\citealt{Geman:1984p721}; \citealt{Press:2007p13558}).}  If the resulting $\chi^2$ value for the trial orbit is less than
the previous $\chi^2$ value, then the trial orbital parameters are retained.  If not, then
the probability of adopting the new value is equal to the ratio of the probabilities from the
previous and trial steps (the Metropolis-Hastings algorithm; \citealt{Metropolis:1953p18293}; \citealt{Hastings:1970p18292}).  If the trial is rejected
then the parameters from the previous step are adopted. 

We altered the standard deviations of the Gaussian transition functions so that the acceptance rates were
between 20\% and 40\% to maximize convergence efficiency.  For each planet-star system,  the initial parameters were
chosen from the best-fitting orbit solutions and each chain was run for 1-3$\times$10$^7$
steps depending on the rate of convergence.  The initial 10\% of each chain was excluded from the final estimation of parameter
uncertainties.  The parameters were tested for convergence by running five shorter chains with 10$^6$ steps. We verified that convergence was reached by ensuring that the
Gelman-Rubin statistic (\citealt{Gelman:1992p18301}; \citealt{Cowles:1996p14537}) for each parameter was near unity (typically $\lesssim$ 1.02) and that a visual inspection of the history plots suggested stability.

\begin{figure}
%     \plotone{f8.eps}
           \resizebox{3.5in}{!}{\includegraphics{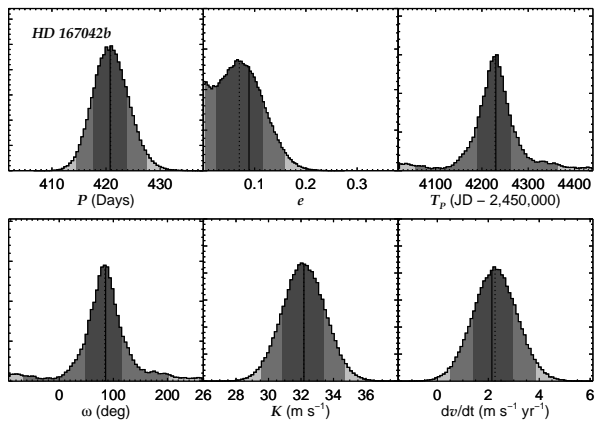}}
  \caption{Markov Chain Monte Carlo posterior probability density distributions for the orbital parameters of HD 167042b.  Median values of the distributions are indicated with dotted lines while the results from the orbit-fitting routine \texttt{RVLIN} are indicated with solid lines.  Dark and medium gray regions display the area encompassing 68.3\% and 95.4\% of the data about the median (corresponding to 1 $\sigma$ and 2 $\sigma$ areas of a Gaussian distribution).   \label{f8.eps} } 
\end{figure}

The results of the MCMC analysis are posterior probability density functions (pdfs) for each orbital parameter used in the model.  An example of the results for HD 167042 is presented in Figure \ref{f8.eps}.  For each parameter we compute the range encompassing 68.3\% of the data about the median of the distribution.  These values represent approximate upper and lower one-sigma limits for Gaussian-like distributions and are included in parentheses next to each parameter in Table \ref{orbitpars1}.  For eccentricities near zero, the 68.3\% range of data about the median is a poor estimator of the most likely values as it fails to encompass the peak of the distribution.  When that occurs we quote upper limit eccentricity values below which encompass 95.4\% of the data.  We note that the results from using the BIC value as a way to discriminate between adopting models with or without velocity trends yields the same results as using the posterior pdf of the velocity trend from MCMC as an indicator.

\begin{figure}
   % \plotone{f9.eps}
      \resizebox{3.5in}{!}{\includegraphics{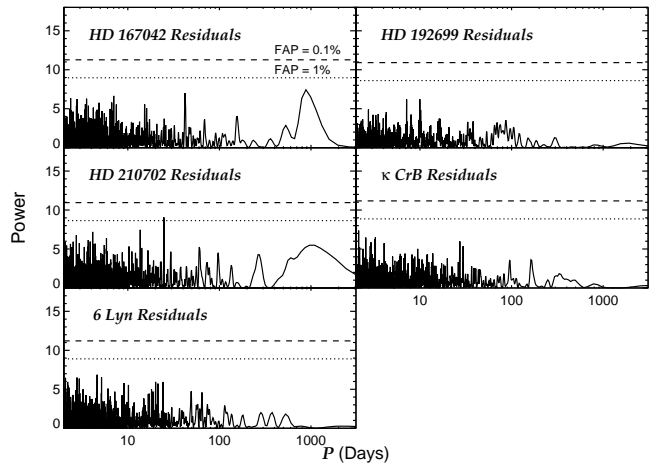}}
  \caption{Lomb-Scargle periodograms for the residuals of five planet-hosting subgiants.  False alarm probabilities (FAP) of 1\% and 0.1\% are shown as dotted and dashed lines, respectively.  No significant periodic signals were detected.      \label{f9.eps} } 
\end{figure}

We use Lomb-Scargle (LS) periodograms (\citealt{Lomb:1976p9259}; \citealt{Scargle:1982p9274}) to search for additional short period companions after subtracting the best-fitting orbit from the observations.  A false alarm probability (FAP) is computed following the method described in \citet{Horne:1986p12729}, which gives the probability that a peak will occur by chance assuming the data are pure noise.  The results are shown in Figure \ref{f9.eps}.  No periodicities in the residuals are identified with FAP of 0.1\% or less.

\subsection{The HD 8375 SB1 System}

\begin{figure}
%    \plotone{f10.eps}
      \resizebox{3.45in}{!}{\includegraphics{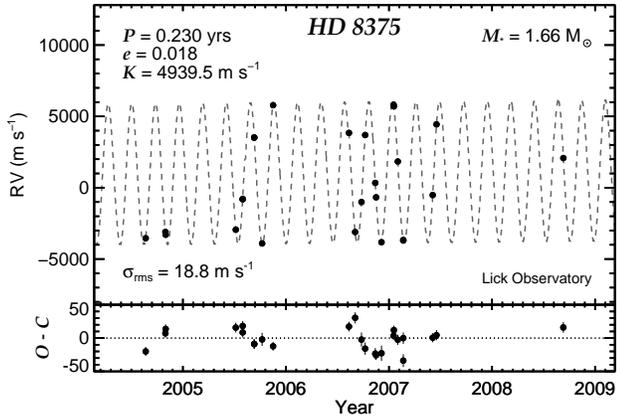}}
  \caption{Lick Observatory radial velocities and best-fitting Keplerian orbital solution for HD 8375. The best-fitting orbit includes the inner companion HD 8375B with a linear velocity trend ($dv/dt$ = 52.9$^{+1.8}_{-1.9}$ m s$^{-1}$ yr$^{-1}$), indicative of an additional long-period companion.  \label{f10.eps} } 
\end{figure}

Several stars in our sample exhibit very high radial velocity scatter indicative of stellar companions orbiting at small semimajor axes.  HD 8375 is one such system and, with 28 observations, is well-enough sampled to derive an accurate orbit solution and posterior pdfs using \texttt{RVLIN} and MCMC (Figure \ref{f10.eps}).  HD 8375 is a known spectroscopic binary (\citealt{Beavers:1986p19149}; \citealt{deMedeiros:1999p19168}; \citealt{Snowden:2005p18744}) but to our knowledge no orbit has been published.  The amplitude and period of the companion suggest that we cannot use the negligible companion mass approximation to derive $M_2$sin$i$.  Instead we compute the mass function $f_1$($m$) and minimum separation $a_1$sin$i$, which are listed along with the orbital parameters of HD 8375B in Table \ref{orbitpars8375}.  We find that a companion with a long-term velocity trend best-fits the data based on BIC values with and without a trend.

\begin{deluxetable}{lc}
\tabletypesize{\scriptsize}
\tablewidth{0pt}
\tablecolumns{2}
\tablecaption{Orbital Parameters for HD 8375B \label{orbitpars8375}}
\tablehead{
    \colhead{Parameter}   &  \colhead{HD 8375B}  }
\startdata
$P$ (days)                                              &   83.9408 (0.0016, 0.0015)  \\
$T_{\mathrm{p}}$ (JD -- 2,450,000)   &   4039.3 (0.5,0.4) \\
$e$                                                          &    0.0179   (0.0004, 0.0004)      \\
$\omega$  ($^{\circ}$)                         &      321.83  (1.74,1.97)   \\
$K$ (m s$^{-1}$)                                   &      4939.2   (2.6,2.5)    \\
$f_1$($m$) (10$^{-3}$ $M_{\odot}$)  &       1.050 (0.002, 0.002)        \\
$a_1$sin$i$ (10$^{-2}$ AU)                &       3.813 (0.002, 0.002)         \\
$\chi^2_{\nu}$                                        &      5.69             \\
BIC (trend)                                              &     \bf{119.3}                     \\
BIC (no trend)                                        &      928.8                     \\
$dv$/$dt$ (m s$^{-1}$ yr$^{-1}$)       &    52.9  (1.8,1.9) \\

\enddata
\tablecomments{See Table \ref{orbitpars1} for details about the orbital parameters and uncertainties.}
\end{deluxetable}

Following \citet{Winn:2009p18838}, we can estimate the minimum mass required to accelerate the HD 8375 system at the observed rate as a function of semimajor axis:

\begin{equation}
M_\mathrm{P}   \sin i \sim \frac{dv}{dt} \frac{a^2}{G} \sec(2\pi \tau),
\end{equation}

\noindent where $M_\mathrm{P}$ is the minimum planet mass, $i$ is the inclination, $dv/dt$ is the acceleration, $G$ is the gravitational constant, and $\tau$ is the ratio of the time baseline of the observations divided by the period of the companion.  For large periods relative to the time baseline, sec(2$\pi \tau$) $\sim$ 1, yielding Equation 1 from \citet{Winn:2009p18838}.  For small values of $\tau$, the measured value of $dv/dt$ gives 

\begin{equation}
M_\mathrm{P} \sin i  \sim 0.295 (a/1 \ \mathrm{AU})^2 M_\mathrm{Jup}.  
\end{equation}

\noindent Values of $M_\mathrm{P}$sin$i$ and $a$ below this curve are not permitted; masses are not high enough at large separations to induce the observed acceleration.  Lowering the value of $\tau$ shifts the curve towards smaller semimajor axes and higher masses.  If the third body is a planet then it is in a circumbinary orbit, otherwise it is a star or brown dwarf in a hierarchical configuration.

\section{Photometry of the Five Known Planetary Systems}

In addition to the radial velocities from Lick Observatory, we used 
the T3 0.4~m and the T12 0.8~m Automated Photometric Telescopes (APTs) at 
Fairborn Observatory to gather time-series photometry of the five systems 
with orbital updates in Table \ref{orbitpars1}.  The observations of these five systems 
were conducted at various epochs between 1993 April and 2009 June.  The 
APTs can detect short-term, low-amplitude brightness variations in cool 
stars due to rotational modulation of the visibility of surface features, 
such as starspots and plages (e.g., \citealt{Henry:1995p19249}), and can also 
detect longer-term variations associated with stellar magnetic cycles 
(e.g., \citealt{Henry:1999p19250}; \citealt{Hall:2009p19260}).  Photometry of planetary candidate host 
stars helps to establish whether observed low-amplitude 
radial velocity variations are caused by stellar activity or 
planetary-reflex motion (e.g., \citealt{Henry:2000p19321}).  \citet{Queloz:2001p19271} and 
\citet{Paulson:2004p3778} have found periodic radial velocity variations in 
solar-type stars that are caused by starspots.  The APT observations are 
also useful to search for possible transits of the planetary companions 
(e.g., \citealt{Henry:2000p19282}; \citealt{Sato:2005p19292}; \citealt{Winn:2009p19320}).

The T12 APT is equipped with a two-channel precision photometer employing 
two EMI 9124QB bi-alkali photomultiplier tubes (PMTs) to make simultaneous 
measurements of a star in Str\"omgren $b$ and $y$ passbands.  The T4 APT 
has a single-channel precision photometer that uses an EMI 9124QB PMT to
measure a star sequentially through $b$ and $y$ filters.  The APTs observe
each target star (star D) in a quartet with three ostensibly constant 
comparison stars (stars A, B, and C).  We compute $b$ and $y$ differential 
magnitudes for each of the six combinations of the four stars: $D-A$, $D-B$, 
$D-C$, $C-A$, $C-B$, and $B-A$.  We then correct the Str\"omgren $b$ and $y$ 
differential magnitudes for differential extinction with nightly extinction 
coefficients and transform them to the Str\"omgren system with yearly mean 
transformation coefficients. Finally, we combine the Str\"omgren $b$ and $y$ 
differential magnitudes into a single $(b+y)/2$ passband to improve the 
precision of the observations.  \citealt{Henry:1999p19250} presents a detailed description
of the automated telescopes and photometers, observing techniques,
and data reduction procedures needed for long-term, high-precision
photometry.

We use the two most constant comparison stars in each quartet to compute
the quantity $\sigma_f$, the target star's absolute night-to-night 
variability, statistically corrected for any unrecognized intrinsic 
variability in the two comparison stars and also corrected for the 
measurement uncertainty of the differential magnitudes.  We follow the 
method described in \citealt{Hall:2009p19260} and compute $\sigma_f$ as 

\begin{equation}
\sigma_{f}^{2} = \sigma_{*}^{2} - \frac{1}{2}\sigma_{c}^{2} - \epsilon^{2},
\end{equation}

\noindent where $\sigma_{*}$ is the standard deviation of the target star's 
differential magnitudes computed as the {\it mean} of the two best $D-A$, 
$D-B$, $D-C$ time series (e.g., $D-(A+B)/2$), $\sigma_{c}$ is the standard 
deviation of the differential magnitudes of the two best comparison stars 
(e.g., $B-A$), and $\epsilon$ is the measurement precision of an individual 
differential magnitude.

We estimate $\epsilon$ by examining the standard deviations of the
comparison star differential magnitudes ($C-A$, $C-B$, and $B-A$) for 
each of the five target stars.  The lowest values are close to 0.0010 mag,
representing the standard deviations of the most stable comparison stars,
which we take to be $\epsilon$. 

The photometric results for the five stars are summarized in Table \ref{photometry}.  
Columns 4 and 5 give the duration of the photometric observations 
in days and the total number of measurements, respectively.  For all stars 
except 6~Lyn, the duration of the photometric measurements approximately
equals or exceeds the orbital period of the planetary companion, though 
not by more than a couple of cycles.  The standard deviations in columns 
6 and 7, $\sigma_*$ and $\sigma_c$, vary somewhat from star to star due to
differences in factors such as stellar brightness, airmass, and seasonal
photometric quality. Our method of computing the absolute variability of 
each target star takes these differences into account.

Finally, the value of each star's absolute variability level, $\sigma_f$, 
is given in column~8 of Table~\ref{photometry}.  All values are significantly less than
0.001 mag (1 mmag) and are consistent with each star being constant.  In 
cases where both the comparison stars and the target star are especially 
stable, random errors in their measurements can result in 
$\sigma_{f}^{2} < 0$, i.e., $\sigma_f$ becomes imaginary.  This is the 
case for $\kappa$~CrB, for which we assume $\sigma_f$ to be zero and 
the star to be constant.  HD~210702 has the largest value of 
$\sigma_f$ (0.00074 mag), though it is still less than 1 mmag.  We classify 
HD~210702 as variable but append a colon to indicate some uncertainty.  
We find nothing significant in our periodogram analyses of all 
five stars.  We conclude that there is no photometric evidence
for brightness variability levels in any of these five stars that could 
call into question the existence of their planetary companions.

\begin{deluxetable*}{ccccccccc}
\tabletypesize{\scriptsize}
%\tablenum{3}
\tablewidth{0pt}
\tablecaption{Photometric Results for the Five Stars in Table \ref{orbitpars1} \newline With Updated Orbital Parameters \label{photometry}}
\tablehead{
\colhead{} & \colhead{} & \colhead{Date Range} & \colhead{Duration} & 
\colhead{} & \colhead{$\sigma_*$} & \colhead{$\sigma_c$} & 
\colhead{$\sigma_f$} & \colhead{} \\
\colhead{Star} & \colhead{APT} & \colhead{(HJD $-$ 2,400,000)} & 
\colhead{(days)} & \colhead{$N_{obs}$} & \colhead{(mag)} & \colhead{(mag)} & 
\colhead{(mag)} & \colhead{Variability} \\
\colhead{(1)} & \colhead{(2)} & \colhead{(3)} & \colhead{(4)} & \colhead{(5)} & 
\colhead{(6)} & \colhead{(7)} & \colhead{(8)} & \colhead{(9)} 
}
\startdata
 HD 167042 & 12 & 54128--55004 &  876 & 275 & 0.00182 & 0.00205 & 0.00046 & Constant  \\
 HD 192699 & 12 & 54192--55004 &  812 & 106 & 0.00214 & 0.00267 & 0.00012 & Constant  \\
 HD 210702 & 12 & 54100--54994 &  894 & 106 & 0.00218 & 0.00253 & 0.00074 & Variable: \\
 $\kappa$ CrB &  4 & 49094--50250 & 1156 & 222 & 0.00094 & 0.00105 & 0.00000 & Constant  \\
   6 Lyn   & 12 & 54437--54928 &  491 & 238 & 0.00153 & 0.00163 & 0.00011 & Constant  \\
\enddata
\end{deluxetable*}

\begin{figure}
    \plotone{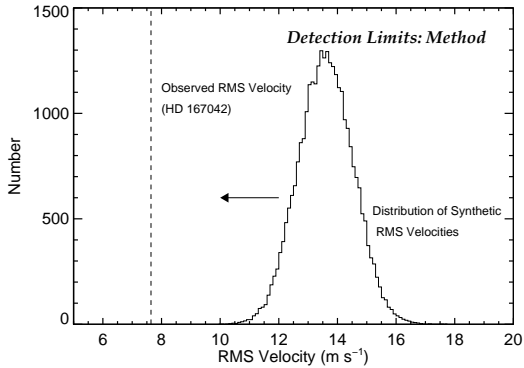}
  \caption{Method for computing detection limits.  For each period we perform a Monte Carlo simulation of radial velocity measurements at the exact dates of the observations beginning with a large velocity semiamplitude and gradually lowering it until the simulated distribution of rms values is consistent with the observed value for that star.  For each period and velocity semiamplitude we test for consistency at some confidence level; in this study we use 95.4\%.  If the observed value is inconsistent with the distribution of rms values, we lower the velocity semiamplitude in the simulations and repeat the process until they are consistent. This particular example is for a trial period of 7.4 days with trial velocity semiamplitude of 16.6 m s $^{-1}$ and 3 $\times$ 10$^4$ trials.  The histogram shows the distribution of rms values derived from our Monte Carlo simulations while the dashed line marks the observed rms value of the residuals of HD 167042.  The distribution of rms values migrates to smaller values as indicated by the arrow.  This process continues for all relevant periods until the detection limits in period-velocity semiamplitude space are derived.    \label{f11.eps} } 
\end{figure}

\section{Detection Limits}

Information about the $lack$ of planets around stars is critical for quantifying the occurrence rates of multiple Jovian planetary systems, computing frequencies of different planet populations (\citealt{Lagrange:2009p15653}), and correcting for detection biases in radial velocity surveys (\citealt{Cumming:2008p9188}).  We use Monte Carlo simulations of radial velocity measurements to assess the detection limits of stars in our sample.  We derive upper limits for the velocity semiamplitude ($K_\mathrm{up}$) as a function of period, which we convert to upper limits for minimum companion mass as a function of semimajor axis using stellar masses.

\subsection{Method}

Our method for computing detection limits is similar to that of \citet{Lagrange:2009p15653}.  We use the rms value of our radial velocities to determine which Keplerian orbits of hypothetical planets are inconsistent with the observed rms values.  For each star we begin with an initial orbital period and a large velocity semiamplitude ($K_\mathrm{trial}$).  We generate a series of synthetic radial velocity measurements with random phases and circular orbits on the same dates that the observations were taken to create a distribution of rms values for that specific period and $K_\mathrm{trial}$.  Noise is added to each synthetic measurement by drawing from a Gaussian distribution with a standard deviation equal to the total measurement-plus-jitter error at that date.  If the distribution is inconsistent with the observed rms value at some confidence level (we chose 95.4\% in this analysis), then $K_\mathrm{trial}$ is decreased and a new distribution is created (Figure \ref{f11.eps}).  If the observed rms value is consistent with the distribution then we adopt that $K_\mathrm{trial}$ value as the velocity semiamplitude upper limit, $K_\mathrm{up}$, for that period.  This process is then repeated over all relevant periods.

\subsection{Subgiants With Known Planets\label{sg_detlimits}}

The detection of a planet around a star increases the probability that other planets are in that system, as finding a planet suggests that the stellar system in question is dynamically and was historically amenable to planet formation and retention. For known planetary systems within 200 pc, \citet{Wright:2009p18840} find that additional planets are present at least 28\% of the time.  It is therefore instructive to analyze the detection limits for the known planet-hosting stars in our sample so that certain regions of mass-semimajor axis phase space may be ruled out as not harboring additional planetary companions.  

\begin{figure}
%     \plotone{f12.eps}
  \resizebox{3.5in}{!}{\includegraphics{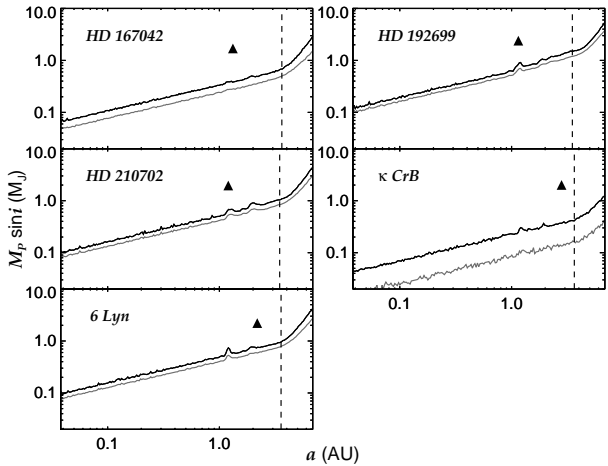}}
  \caption{Detection limits for five planet-hosting subgiants based on Monte Carlo simulations of radial velocity measurements.  The solid black and gray curves show the 95.4\% and 68.3\% minimum mass upper limit as a function of semimajor axis.  Triangles show the locations of the known planets in these systems.  The dashed lines indicate the semimajor axes that correspond to the baseline of the observations for that particular target.  Planets located at larger semimajor axes will have completed less than one full period from the first to the last radial velocity measurement, hence the rise in the detection limit curves in that region.  \label{f12.eps} } 
\end{figure}

To derive upper limits we generate 2500 artificial Keplerian orbits for each $K_\mathrm{trial}$.  The $K_\mathrm{trial}$ value is lowered in steps of 0.1 m s$^{-1}$ until the rms distribution is consistent with the observed rms value at the 95.4\% level.  200 periods are evenly sampled from a log($P$) distribution between 2 to 5000 days.  Results for the five IM subgiants are presented in Figure \ref{f12.eps} and Table \ref{detlimits}.

Our observations are sensitive to planet minimum masses between a few times that of Neptune (0.053 $M_\mathrm{Jup}$) to several Jupiter masses for semimajor axes between 0.04 AU to several AU.  With a high probability we are able to exclude the existence of additional planets in the upper portions of each plot in Figure \ref{f12.eps}.  However, Neptune-mass planets in circular orbits cannot be ruled out by our observations.  In general our detection limits are sensitive to the observed rms value, the internal measurement errors, and the assumed stellar jitter level.  It is therefore unlikely that more observations will significantly improve the resulting detection limits as stellar jitter is the dominant contribution to the rms budget.  IM main sequence stars and giants both have higher mean jitter levels than subgiants (e.g., \citealt{Lagrange:2009p15653}; \citealt{Setiawan:2003p10012}); transit or microlensing surveys may therefore be the only tools available to study sub-Neptune-mass planets in circular orbits around IM stars, although the larger radii and shorter evolutionary lifetimes of IM stars will make the detection of low-mass planets difficult even using these techniques.

\subsection{Subgiants from the Lick Survey}

Overall our sample contains a variety of radial velocity characteristics (see the ``Notes'' column of Table \ref{detlimits} for a summary): five targets are known planet-hosting stars, nine exhibit linear velocity trends, and four have large rms velocity scatter $>$ 900 m s$^{-1}$ (which roughly corresponds to the semiamplitude of a 13 $M_\mathrm{Jup}$ object orbiting a 1.5 $M_{\sun}$ star at 0.1 AU).  The targets with large velocity scatter are all known spectroscopic binaries and include HD 8375 (discussed in \S3.1), HD 65938 (\citealt{Pourbaix:2004p19187}; \citealt{Massarotti:2008p19189}), HD 179799 (\citealt{Setiawan:2003p10012}; \citealt{Setiawan:2004p19094}; \citealt{Massarotti:2008p19189}), and HD 210211 (spectroscopic triple system; \citealt{Horch:2002p19200}; \citealt{Pourbaix:2004p19187}; \citealt{Massarotti:2008p19189}).

The two systems HD 90043 and HD 200964 both exhibit long-period ($\sim$ 1.5 yr) radial velocity variations which are likely caused by planetary-mass companions.  Both stars are chromospherically quiet and do not show strong photometric variability.\footnote{Time-series photometric observations will be presented with the orbital solutions of HD 90043 and HD 200964 in a future publication.}  LS periodograms confirm strong periodicities for both systems with false alarm probabilities $<$ 0.1\%.  Our radial velocity coverage is currently insufficient to obtain unique orbital solutions so we withhold a detailed discussion of these objects for a future publication.  However, we note that single-planet orbital solutions fail to accurately reproduce the observed variations; instead, two-planet solutions produce significantly better fits.  The inner planet of HD 90043 has a minimum mass of $\sim$ 2.5 $M_\mathrm{Jup}$ and is located at $\sim$ 1.4 AU ($P$ $\sim$ 1.2 yr), while the inner planet of HD 200964 has a minimum mass of $\sim$ 2.0 $M_\mathrm{Jup}$ and is located at $\sim$ 1.7 AU ($P$ $\sim$ 1.7 yr).  No additional stars in our sample exhibit periodicities with FAP $<$ 0.1\%.

We derive detection limits for our full sample using the same parameters as in $\S$\ref{sg_detlimits} for all stars with more than five radial velocity measurements.  The results are presented in Table \ref{detlimits}.  We subtract the best-fitting linear trends for stars exhibiting constant accelerations.  Excluding the large rms velocity stars, the median (minimum) mass upper limits for our sample are $\sim$ \{0.21, 0.34, 0.50, 0.64, 1.3\} $M_\mathrm{Jup}$ at semimajor axes of \{0.1, 0.3, 0.6, 1.0, 3.0\} AU at the 95.4\% confidence level.  Our sample of subgiants is therefore typically sensitive to Saturn-mass planets out to $\sim$ 0.1 AU and Jupiter-mass planets out to $\sim$ 1 AU.  Excluding stars with large rms velocity values, planets $>$ 3 $M_\mathrm{Jup}$ can be excluded out to 1 AU for all stars in our sample, and planets $>$ 6 $M_\mathrm{Jup}$  can be excluded out to 3 AU for all but one of our stars.  For the majority of our targets, however, our survey is sensitive to a few Jupiter-mass planets within 3 AU, or 1 $M_\mathrm{Jup}$ planets within a few AU.

\begin{deluxetable*}{lccccccccccccc}
%\rotate
\tabletypesize{\scriptsize}
\tablewidth{0pt}
\tablecolumns{14}
\tablecaption{Detection Limits \label{detlimits}}
\tablehead{
\colhead{} & \colhead{}   &  \colhead{Other} &  \colhead{$M_*$} & \multicolumn{5}{c}{$M_P$ sin$i$ ($M_\mathrm{Jup}$) for $a$ } &  \colhead{rms} & \colhead{}  &  \colhead{Baseline} &  \colhead{Trend}  &  \colhead{}     \\
\cline{5-9}
        \colhead{HD}  & \colhead{HIP}  & \colhead{Name}  & \colhead{($M_{\sun}$)}  &  \colhead{$<$ 0.1 AU}    & \colhead{$<$ 0.3 AU}  & \colhead{$<$ 0.6 AU} &  \colhead{$<$ 1 AU} &  \colhead{$<$ 3 AU} & \colhead{(m s$^{-1}$)} & \colhead{N$_\mathrm{obs}$} &  \colhead{(yr)} & \colhead{(m s$^{-1}$ yr$^{-1}$)} &  \colhead{Notes\tablenotemark{a}} 
}
\startdata
\cutinhead{Planet-Hosting Subgiants (Residuals)}
45410      &   31039  &   6 Lyn    & 1.82  &  0.15   &     0.27  &  0.37  &   0.49  &   0.86   & 9.3    & 64  &   5.01     &  \nodata &    P  \\
90043      &  50887  &   24 Sex    & 1.91  &  0.12   &     0.20   &  0.29     &   0.37 &   0.73   & 8.4  & 50 & 4.37    & \nodata &   PP   \\     
142091    &  77655  &   $\kappa$ CrB    & 1.84  &  0.07   &     0.12  &  0.17      &   0.22  &   0.37   & 6.2  &  62 &  5.15    & \nodata &  P  \\ 
167042    &  89047     &    \nodata    & 1.72 &  0.11   &     0.18  &   0.25     &   0.34  &   0.59   &  7.7  &  68 & 5.27    &  2.1 & P, LT  \\ 
192699    &   99894    &   \nodata     & 1.69 &  0.20   &     0.34   &  0.49     &   0.62  &   1.40   & 10.5 & 50  &  5.05  &  \nodata  &  P   \\ 
200964    &   104202  &   \nodata    &  1.70 &  0.13   &     0.22   &   0.33    &   0.42  &   0.81   & 8.7 & 91  & 5.02    &  \nodata   & PP   \\ 
210702    &  109577   &    \nodata    & 1.85  &  0.17   &     0.32   & 0.39    &   0.51  &   0.98   & 8.8  & 51   & 4.76      &  \nodata &   P   \\ 
 \cutinhead{Lick Observatory Subgiant Sample}
587        &  840    &    \nodata  & 1.72    &     0.20   &     0.33  &  0.42      &   0.61  &   1.05    & 8.17       & 22 & 3.90   &  --5.9  &  LT    \\ 
8110     &   6289 &   \nodata  & 1.64   &     0.56   &     1.04  &  1.41      &   2.13  &   5.35    & 21.3     & 10  & 2.02    & \nodata  &    \\ 
8375\tablenotemark{b}   &  6512  &   \nodata  &  1.66   &     0.42    &     0.68 &   0.96      &   1.25   &  2.49    & 19.0    &  25  & 4.05  &  52.9  &   SB, LT      \\ 
22682   &   17049 &   \nodata  & 1.65  &     0.34   &     0.71  &  0.88      &   1.19  &   1.98    & 13.2    & 10 & 3.85    &   \nodata    &   \\ 
25975   &  19302  &  \nodata   &  1.59 &     0.21   &     0.32 &   0.44      &   0.63  &   0.95    & 7.71    & 16 & 4.69   &    \nodata  &   \\ 
27536   &  20263  & EK Eri   &  1.87 &     0.77   &     1.36   & 2.09      &   2.95  &   5.06    & 32.6    & 36  & 3.24     &  \nodata &    \\ 
34538   &  24679  &  \nodata   & 1.53  &     0.28   &     0.51  &  0.70      &   0.86  &   1.66    & 11.2   & 12 &  4.13 & \nodata  &      \\ 
37601   & 26942   &  24 Cam  &  1.93 &     0.28   &     0.42  &  0.68      &   0.86  &   1.54    & 10.1   & 17 & 4.01  &   9.2      &    LT   \\ 
45506   &   30815  &   \nodata  & 1.84  &     0.16   &     0.29 &   0.43      &   0.53  &   0.84    & 6.80   & 14 & 4.21 &   --21   &   LT   \\ 
57707   &  35751  &   \nodata  &  1.81 &     0.40   &     0.91 &   0.95      &   1.17  &   2.25    & 10.8   & 6  & 2.24  &   \nodata &    \\
65938   &  39198  &   \nodata  &  1.66  & \nodata  & \nodata &  \nodata   & \nodata  & \nodata & 2060 &  4  & 0.99  & \nodata &   SB      \\ 
73764   &  42528  &   \nodata  &  1.91 &     0.48   &     0.73 &   0.96      &   1.33  &   1.99    & 13.1      & 9  &  3.30 & \nodata  &   \\ 
103484  &  58110  &  6 Vir  &  1.98 &     0.70   &     1.10  & 1.62       &   2.71  &   3.90    & 25.8     &  40 & 4.09 &  \nodata &    \\ 
109272  &  61296  &  \nodata   & 1.83  &     0.22   &     0.62  & 0.70       &   0.73  &   1.26    & 8.81     &  16 & 3.87 &  \nodata  &    \\ 
111028  & 62325   &  33 Vir  & 1.69  &     0.19   &     0.34   & 0.47       &   0.60  &   1.07    & 9.28    & 23 & 4.03  &  --9.9  &  LT  \\ 
123929  & 69185   &    \nodata & 1.61  &     0.50   &     0.59   &1.56       &   2.40  &   2.19    & 12.3     & 7 &  2.62   & \nodata   &    \\ 
135944  &  74690  &  \nodata   & 1.62  &     0.17   &     0.31  &  0.43       &   0.52  &   0.95    & 7.68     & 16 & 4.10 &  53  &  LT    \\ 
153226  &  82989  &  \nodata   &  1.76 &     0.19   &     0.33   &  0.50     &   0.64  &   1.04    & 7.85     & 18  & 3.93  &  \nodata &  \\ 
179799  &  94521  &   \nodata  &  1.74 &      42.8   &    85.0   &  129     &  177     &   355     & 2510   &  8  & 3.67 &   \nodata  &   SB      \\ 
184010  &  96016  &   \nodata  & 1.80  &     0.37   &     0.64  &   0.93     &   1.13  &   2.35    & 15.2     &  26 & 4.07 &  \nodata  &   \\ 
185351  &  96459  &  \nodata   &   1.91 &   0.17     &    0.31  &   0.43    &  0.52     &   1.11  & 7.09    &  17  & 3.93 &  \nodata    &         \\ 
202568  & 104941   &  \nodata   & 1.57  &     0.20   &     0.31  &   0.40     &   0.54  &   1.04    & 7.10        &  11 & 3.82 &  --5.2  &  LT   \\ 
210211  & 109281   &  \nodata   & 1.82  &    125    &       228  &   330     &    438    &   794    & 7930   &  8  & 2.25 &  \nodata     &   SB      \\ 
210404  &  109338  &  \nodata   &  1.75 &     0.68   &     1.21  &  1.77      &   2.44  &   14.8    & 26.3      & 11 & 1.21 &  --209  &   LT   \\ 
\enddata
\tablenotetext{a}{``P'' = planet; ``LT'' = linear trend;  ``PP'' = probable planet, ``SB'' = spectroscopic binary (or higher order system).}
\tablenotetext{b}{Detection limits are for the residuals after subtracting off the stellar companion and the linear velocity trend.}
\tablecomments{Detection limits are for 95.4\% confidence levels.  Linear trends have been subtracted from targets with ``LT.''}
\end{deluxetable*}

\section{Jovian Planet Frequency Around Intermediate-Mass Stars}

Several studies have provided hints that the occurrence rate of Jovian planets increases with stellar mass.  \citet{Johnson:2007p169} combine the results of the California and Carnegie Planet Search around FGK and M-type stars with the frequency of planets using the entire Lick IM Subgiant Sample.  They found that the frequency of Jovian planets appears to rise with stellar mass, reaching estimated rates of at least 9\% for IM stars.  Similarly, \citet{Lovis:2007p17712} found a higher frequency of massive planets ($M_\mathrm{P}$ sin$i$ $>$ 5 $M_\mathrm{Jup}$) around IM stars than around solar-type stars.  These previous estimates of Jovian planet frequency, however, have combined multiple published samples with different radial velocity sensitivities and different sample selection criteria.  While our sample size is considerably smaller in this work, we benefit from having a uniform data set observed over a long ($\sim$ 5-year) baseline with the same telescope and instrument setup.  We also have the advantage of having well-characterized detection limits for the entire sample.

Five stars in our sample are known planet-hosting subgiants and we identify two more (HD 200964 and HD 90043) as being strong candidates for having planetary companions ($\S$5.3).  The number of Jovian planet-hosting stars in our sample of 31 subgiants is therefore at least 7.  Bayes' Theorem can be used to derive the posterior pdf for the frequency of planets within $\sim$ 3 AU:

\begin{equation}
P(p \vert k, n) \propto P(k \vert p, n) P(p),
\end{equation}

\noindent where $P(p \vert k, n)$ is the posterior pdf for the probability of a star hosting a planet ($p$) given $k$ detections in a sample of $n$ stars, $P(k \vert p, n)$ is the pdf for observing $k$ detections in the sample, and $P(p)$ is the prior pdf for the probability that a star hosts a planet.  The detection of a planet represents a Bernoulli trial, so $P(k \vert p, n)$ follows a binomial distribution for the unknown parameter $p$.  If we assume a uniform prior for $P(p)$ between 0 and 1, the posterior pdf $P(p \vert k, n)$ simply follows a binomial distribution.

\begin{figure}
 %    \plotone{f13.eps}
       \resizebox{3.5in}{!}{\includegraphics{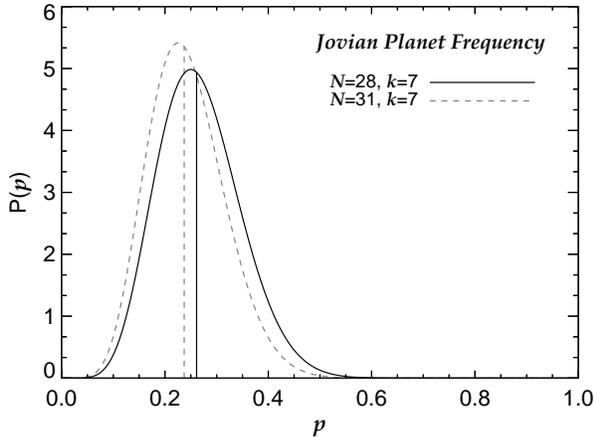}}
  \caption{Probability density function ($P(p)$) for the frequency of Jovian-mass planets ($p$) around intermediate-mass stars interior to $\sim$ 3 AU. Assuming a uniform prior, the posterior pdf for the probability of harboring a planet follows a binomial distribution for $k$ ``successes'' (planets detected) out of a sample size of $N$ targets.  The dashed gray curve shows the distribution for the entire sample ($N$ = 31 intermediate-mass subgiants), while the dashed gray line indicates the median of the distribution (24\%).  Several targets exhibited large radial velocity variations indicating a close stellar or brown dwarf companion; the black curve shows the posterior pdf excluding subgiants with large rms velocity scatter (28).  The median value of this distribution is 26\%.  \label{f13.eps} } 
\end{figure}

The resulting posterior pdfs are shown in Figure \ref{f13.eps}.  The median fraction of planet-hosting subgiants in our sample is 24$^{+8}_{-7}$\%, with upper and lower limits representing the range encompassing 68.3\% of the distribution about the median.  If we exclude subgiants exhibiting large rms velocity scatter, which is indicative of a stellar companion, then the fraction increases to 26$^{+9}_{-8}$\%.    These values can be compared to quantitative predictions from theoretical modeling of planet formation.  For the mass range of our Lick subgiant sample (1.5-2.0 $M_{\sun}$), \citet{Kennedy:2008p18349} predict Jovian planet occurrence rates to be $\sim$ 10-15\%.    Our observed frequency is significantly higher, which may suggest that planet formation around IM stars is more efficient than previously thought.  The lower range of our results are, however, marginally consistent with the upper range predicted by \citet{Kennedy:2008p18349}.

Our results can be compared to the giant planet frequencies derived by other authors for solar-type stars.  Our survey is sensitive to giant planets with masses $\gtrsim$ 1 $M_\mathrm{Jup}$ within a few AU.   \citet{Johnson:2007p169} derive a planet frequency of $\sim$ 4\% for solar-type stars covering similar planet mass and semimajor axis ranges; the giant planet frequency they derive is significantly lower than the frequency we find for IM stars.  In a similar analysis, \citet{Lovis:2007p17712} derive a frequency of $\sim$0.5\% for planets with masses $>$ 5 $M_\mathrm{Jup}$ within 2.5 AU for solar-type stars.  Over the same planet mass and semimajor axis range our program did not detect any planets. Using the binomial theorem, zero out of 28 stars translates into a 2 $\sigma$ upper limit of 10.1\% for the frequency of $>$ 5 $M_\mathrm{Jup}$ planets within 2.5 AU.  Although poorly constrained, our frequency of the most massive giant planets around IM stars is consistent with the results from \citet{Lovis:2007p17712} for solar-type stars.

Finally, we note that, intriguingly, the planet occurrence rates for solar- and intermediate-mass stars resemble the occurrence rates of debris disks found through excess infrared emission ($\sim$15\% for Sun-like stars: \citealt{Bryden:2006p19230}; \citealt{Trilling:2008p19220}; $\gtrsim$30\% for IM stars: \citealt{Rieke:2005p18823}; \citealt{Su:2006p19210}; \citealt{Morales:2009p18785}), although at the moment no correlation has been found between the presence of planets and debris disks (\citealt{MoroMartin:2007p19678}; \citealt{Kospal:2009p18893}; \citealt{Bryden:2009p19677}).

\section{The Mass-Period Distribution and Planet Population Synthesis}

The mass-period distribution of extrasolar planets around solar-type stars has been extensively studied in the literature (e.g., \citealt{Tabachnik:2002p18829};  \citealt{Lineweaver:2003p18830}; \citealt{Cumming:2008p9188}).  The simplest and most common parametric technique to model the distribution of giant planets is to fit a double power law to a mass and period histogram:

\begin{equation} \label{eqn:mp}
dN \propto M^{\alpha} P^{\beta} d \ln M d \ln P,
\end{equation}

\noindent where $dN$ is the number density of objects, $M$ is the planet (minimum) mass, and $P$ is the planet's orbital period.  Values of $\alpha$ and $\beta$ are generally consistent among authors; we take the values from \citet{Cumming:2008p9188} as being representative of studies in the literature for Sun-like stars.  Cumming et al.  find $\alpha$ = --0.31 $\pm$ 0.2 and $\beta$ = 0.26 $\pm$ 0.1 for masses $>$ 0.3 $M_\mathrm{Jup}$ and periods $<$ 2000 days.  Qualitatively this means that the mass and period distributions for planets around solar-type stars rise sharply (in linear space) for lower masses and shorter periods, respectively.\footnote{Equation \ref{eqn:mp} can also be expressed as $dN$ $\propto$ $M^{\alpha - 1} P^{\beta - 1} dM dP$.  In logarithmically-spaced bins  the number of planets remains nearly constant.}  The probability of finding a low-mass, short-period planet is therefore much higher than finding a high-mass, long-period planet.  The masses and periods of planets being discovered around IM stars is qualitatively quite different from those being discovered around solar-type stars.  For example, out of $>$ 20 planets discovered around IM stars, none orbit at semimajor axes less than 0.6 AU  ($\sim$ 130 days for a 1.75 $M_{\odot}$ star) and none have minimum masses below 0.70 $M_\mathrm{Jup}$ (see Figure \ref{f1.eps} and Table \ref{evolvedtab}).  This has generally been attributed to a different mass-period distribution for planets around IM stars, but the influence of higher jitter levels on detection limits in surveys targeting IM stars is usually not taken into account and no quantitative analysis has yet been performed.  

We use a Monte Carlo method to rigorously test the null hypothesis that the mass-period distribution for planets around solar-type and IM stars is the same.  The question we seek to answer is the following: Given the values of $\alpha$ and $\beta$ for solar-type stars, a planet occurrence rate, and the detection limits for our sample of 28 stars,\footnote{Three targets (HD 65938, HD 179799, and HD 210211) exhibit large rms velocity scatter and have either too few radial velocity measurements or too short time baselines to accurately fit orbit solutions, so the detection limits of the residuals are not available.  We therefore exclude these targets from the simulation.} what is the probability of finding 7 or more planets all having masses greater than 1.5 $M_\mathrm{Jup}$ and semimajor axes greater than 1 AU?\footnote{In hypothesis testing, the probability of choosing extreme values (tail integrals) must be used rather than the probability of choosing a specific value.  One could easily conceive of a discrete distribution in which the probability of choosing any $particular$ value is small, so in this study we derive the probability of finding 7 \emph{or more} planets with masses \emph{greater than} 1.5 $M_\mathrm{Jup}$ and semimajor axes \emph{greater than} 1 AU.}

We first test whether we can reproduce the observed number and properties of planets in our sample using the same Jovian planet frequency observed around solar-type stars.  For each trial we randomly draw a new value of $\alpha$ and $\beta$ from a Gaussian distribution centered on the values from Cumming et al. with a standard deviation equal to the quoted uncertainties.  For each star in our sample we use the 10.5\% planet frequency derived by  Cumming et al.  to determine whether that particular star harbors a planet in our simulation.  If a star harbors a planet, we randomly draw a mass and period for that planet from the power law distribution in Equation \ref{eqn:mp}.  Finally, we check to see whether that planet would have been detected around that star based on the star's detection limits.  We repeat this process $10^5$ times, saving the results after each trial.

\begin{figure}
    % \plotone{f14.eps}
      \resizebox{3.5in}{!}{\includegraphics{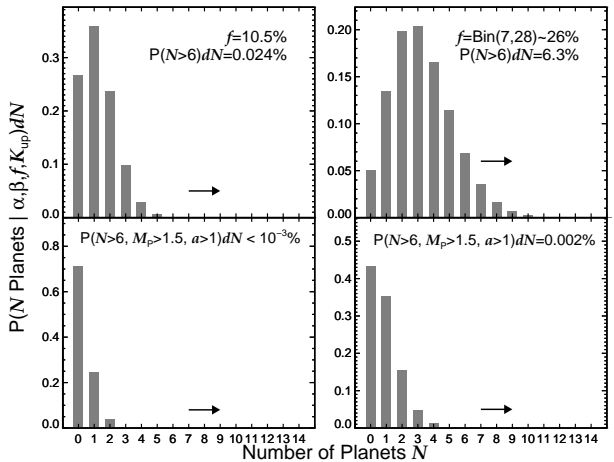}}
  \caption{Comparison of the mass-period distributions of solar-type stars to intermediate-mass stars.  Each histogram displays the results of a Monte Carlo planet population simulation based on values of $\alpha$ and $\beta$ from \citet{Cumming:2008p9188} following the power law distribution in Equation \ref{eqn:mp}, an input planet frequency, and the detection limits from our sample.  Using the planet frequency around solar-type stars (10.5\%), the probability of detecting 7 or more planets is 0.024\% (top left), while the probability of detecting $\ge$ 7 planets with masses $>$ 1.5 $M_\mathrm{Jup}$ and semimajor axes $>$ 1 AU is less than 10$^{-3}$\% (bottom left).  When we adjust the input planet frequency so that each trial draws a new frequency value from a binomial distribution with 7 successes out of a sample of 31, 6.3\% of the trials yield $\ge$ 7 planets (top right), but only 0.002\% of the trials yield $\ge$ 7 planets with masses $>$ 1.5 $M_\mathrm{Jup}$ and semimajor axes $>$ 1 AU (bottom right).   \label{f14.eps} } 
\end{figure}

The results of our first Monte Carlo simulation are shown in the left panels of Figure \ref{f14.eps}.  Out of $10^5$ trials, 24 produced 7 or more planets with any mass or semimajor axis, yielding a probability of 0.024\% (Figure \ref{f14.eps}: top left).  From the same simulation, however, no trials produced 7 or more planets \emph{with masses above 1.5 $M_\mathrm{Jup}$ and semimajor axes greater than 1 AU}, yielding a probability of less than 10$^{-5}$ (Figure \ref{f14.eps}: bottom left).  The input model values of $\alpha$ and $\beta$ and/or the input planet frequency are therefore inconsistent with the number and properties of planets we observed in our sample of intermediate-mass stars.

To determine whether increasing the planet frequency changes the results, we perform a similar Monte Carlo simulation except instead of the 10.5\% occurrence rate we randomly draw a new planet frequency for each trial following the binomial distribution for 7 detections out of a sample of 28 (yielding a median planet frequency of $\sim$ 26\%).  For this exercise, the values of $\alpha$ and $\beta$ are the same as the Cumming et al. values, but the probability of a star hosting a planet changes for each trial.  The results of this simulation  are presented in the right-hand panels of Figure \ref{f14.eps}.  The probability of detecting 7 or more planets with any mass and semimajor axis rose to 6.3\%, but the probability of detecting 7 or more planets with masses above 1.5 $M_\mathrm{Jup}$ and semimajor axes larger than 1 AU is 0.002\%.  Even when accounting for the different planet occurrence rates, we can rule out the Sun-like mass-period distribution for planets around IM stars at a confidence level of over 4 $\sigma$.

\begin{figure*}
    \plotone{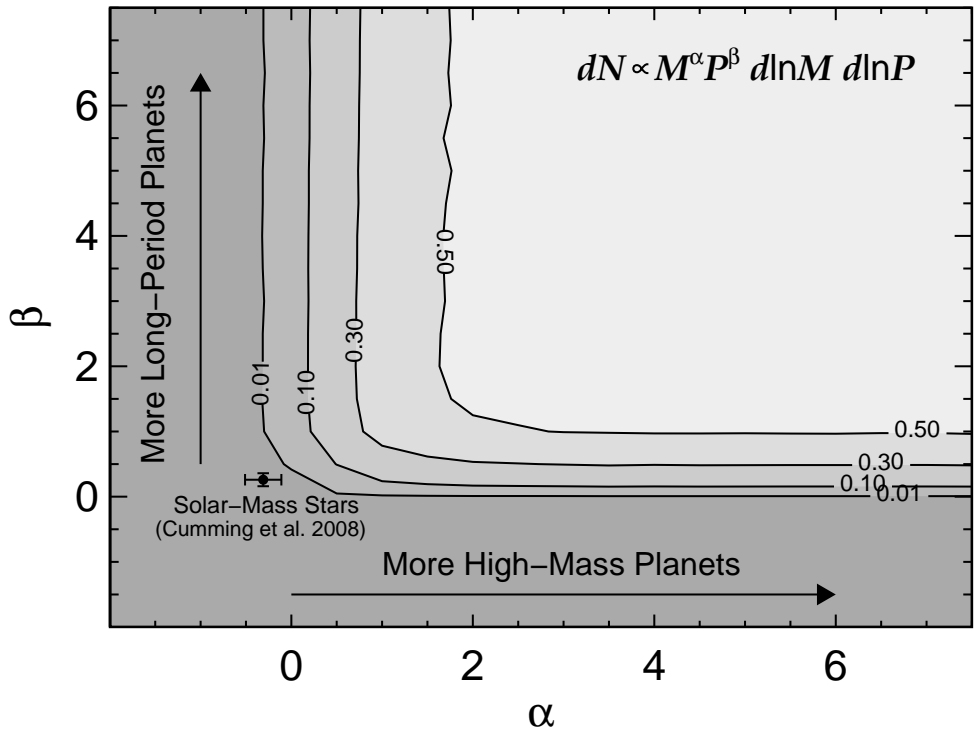}
 %      \resizebox{3.45in}{!}{\includegraphics{f15.eps}}
  \caption{Confidence regions for excluding pairs of $\alpha$ and $\beta$ exponents for the parametric mass-period distribution $dN$ $\propto$ $M^{\alpha}P^{\beta}$$d$ln$M$$d$ln$P$ based on planets in our sample.  The values at each $\alpha$-$\beta$ pair indicate the fraction of the time our Monte Carlo simulations yield the high number ($\ge$ 7) and the properties ($M_\mathrm{P}$ $>$ 1.5 $M_\mathrm{Jup}$, $a$ $>$ 1 AU) of planets observed in our sample of intermediate-mass stars.  We can exclude small values of $\alpha$ and $\beta$ with a high level of confidence.  The mass-period distribution for Sun-like stars from \citet{Cumming:2008p9188} is labeled.  \label{f15.eps} } 
\end{figure*}

These simulations can easily be extended to exclude a wider range of $\alpha$ and $\beta$ parameter space with varying degrees of confidence.  We do this by running the same Monte Carlo simulation drawing from a binomial planet frequency distribution (with 7 detections out of 28 stars) over the following ranges of exponents: --2 $<$ $\alpha$ $<$ 8 ($\Delta \alpha$ = 0.5) and --2 $<$ $\beta$ $<$ 8 ($\Delta \beta$ = 0.5).  For each $\alpha$-$\beta$ pair we run $10^3$ trials.  The results are displayed in Figure \ref{f15.eps}; each value of the grid represents the probability of detecting 7 or more planets with masses above 1.5 $M_\mathrm{Jup}$ and semimajor axes greater than 1 AU.  This technique allows us to exclude large regions of $\alpha$-$\beta$ space as being inconsistent with our observations.  Negative values and small positive values of $\alpha$ and $\beta$ are rejected with high confidence levels, while larger values of $\alpha$ and $\beta$ are able to reproduce the observed properties more often.  Qualitatively this makes sense: we did not detect any low-mass (small $\alpha$) or short-period (small $\beta$) planets.  We note that the inclusion of detection limits is critical in our analysis as it allows us to ``detect'' or ''miss'' simulated planets based on whether they fall above or below our detection threshold.

\section{Discussion}

Little is known about the Saturn- or Neptune-mass population of planets around IM stars.  Most planets orbiting IM stars discovered to date have masses $\gtrsim$ 2 $M_\mathrm{Jup}$ (Table \ref{evolvedtab}), leading to suggestions that more massive stars tend to produce more massive planets (\citealt{Lovis:2007p17712}).  It is unclear, however, whether this is a \emph{bona fide} property or a result of an observational bias caused by higher jitter levels in IM stars which might mask the signals of lower-mass planets.  Moreover, few authors have assessed the limiting planet masses attainable by Doppler surveys of evolved IM stars.  Despite the dearth of Jovian planets at small semimajor axes, could a population of Saturn-mass planets exist interior to 1 AU around  IM stars?  Would the higher jitter levels of evolved IM stars prevent the discovery of Neptune-mass planets?  We address these questions in this study by deriving detection limits for our sample of 31 IM (1.5 $<$ $M_*/M_{\odot}$ $<$ 2.0) subgiants.

Typical detection limits for our stellar sample include planet masses down to $\sim$ \{0.21, 0.34, 0.50, 0.64, 1.3\} $M_\mathrm{Jup}$ within \{0.1, 0.3, 0.6, 1.0, 3.0\} AU at the 95.4\% confidence level, excluding targets with large radial velocity scatter.  We can therefore rule out the existence of hot Saturns within 0.1 AU and Jovian planets out to 1 AU for most stars in our sample.  These detection limits suggest that the notably high masses of planets from our sample ($\gtrsim$ 2 $M_\mathrm{Jup}$) compared to planet masses around solar-type stars may be caused by a real difference in planet population characteristics.  For example, if a population of planets with random masses existed at 1 AU then we would expect an observational bias to result in an observed mass distribution that was truncated near the typical mass detection limit.  As Jovian planets would have been detected out to $\sim$ 1 AU, the higher planet masses uncovered so far may be indicative of a real trend.

We test this idea quantitatively by comparing the mass-period distribution of planets around Sun-like stars to the observed number and properties of planets in our sample.  Even when correcting for the higher planet occurrence rate found in our sample, the values of $\alpha$ and $\beta$ in Equation \ref{eqn:mp} fail to reproduce the number (7/28), masses ($>$ 1.5 $M_\mathrm{Jup}$), and semimajor axes ($>$ 1 AU) of planets from our sample at a confidence level of $>$ 4 $\sigma$.  We conclude that the frequency and mass-period distribution of planets around IM stars is different from those around solar-type stars.  Increasing the mass of the host star by a mere factor of 1.5-2 results in an entirely new planet population which is characterized by a high frequency ($\sim$ 26\%) of high-mass planets ($M_\mathrm{P}$sin$i$ $>$ 1.5 $M_\mathrm{Jup}$) at large semimajor axes ($a$ $>$ 1 AU).

The detection limits of the residuals of the planet-hosting stars in our sample show the strength of subgiant jitter levels and therefore trace the sensitivity levels attainable to Doppler surveys targeting subgiants.  The residual rms velocities range from $\sim$6-10 m s$^{-1}$.  The detection limits indicate that  planetary companions with minimum masses above \{0.2, 0.4, 0.5, 0.7, 1.4\} $M_\mathrm{Jup}$ within \{0.1, 0.3, 0.6, 1.0, 3.0\} AU cannot exist in these systems.  If these detection limits are representative of the typical jitter levels of subgiants then the higher rms velocity values observed in other subgiants from our sample may suggest as-yet-unrecognized low-mass companions.  Given our jitter-dominated detection limits, the prospects of discovering Neptune-mass planets ($\sim$ 0.053 $M_\mathrm{Jup}$) in circular orbits around IM stars using the Doppler technique is not encouraging.  If the dominant source of jitter in subgiants is from $p$-mode oscillations, observing strategies that include longer integration times or repeated exposures over hour-long timescales may help to partially overcome this hurtle (see \citealt{Otoole:2008p9318}).  Our ongoing survey of intermediate-mass subgiants at Keck Observatory will address this issue by using the high RV precision achievable with the HIRES spectrometer and larger telescope aperture of Keck compared to the Lick 3m.

We derive updated orbit solutions using new observations for five of the known planet-hosting IM subgiants in our sample, which were originally announced by \citet{Johnson:2007p165}, \citet{Johnson:2008p166}, and \citet{Sato:2008p14857}.  Our results are in excellent agreement with those previously reported in the literature.  Our parameter uncertainties are typically a factor of $\sim$ 2 smaller than the published values as a result of longer baselines and more radial velocity measurements.  We also note that there is no evidence for periodicity in the residuals of the previously-known planet-hosting stars. We acquired time-series photometric observations of the five known planet-hosting stars with the Automatic Photometric Telescopes at Fairborn Observatory.  We find no evidence for brightness variation levels in any 
of the five stars that could call the existence of their planetary companions into question.  

The eccentricities of three of the five planets in our sample with accurate orbital solutions are consistent with zero and emphasize a low-eccentricity trend for planets around IM subgiants, all but two of which have eccentricities below 0.3 (Table \ref{evolvedtab}).  This is in contrast to planets around solar-type stars, which have approximately uniform eccentricity distributions between 0.0 and 0.8 for semimajor axes $\gtrsim$ 0.3 AU (\citealt{Butler:2006p3743}; \citealt{Wright:2009p18840}).  Interestingly, \citet{Wright:2009p18840} find that the eccentricity distributions for planets with masses $<$ 1 $M_\mathrm{Jup}$ peaks at $e$ $<$ 0.2, while the planets with masses $>$ 1 $M_\mathrm{Jup}$ have more uniformly distributed eccentricities between 0.0 $<$ $e$ $<$ 0.6.  Planets around IM stars show the opposite trend: high-mass planets tend to have circular orbits.  We will explore this effect in detail in a future publication in this series.

Information about planets orbiting IM stars has been driven by observations rather than theory.  There is a growing need for stellar mass-dependent theoretical models of planet formation that make testable predictions about the physical and orbital characteristics of planets in this stellar mass regime.  Specifically, the abundance of low-mass planets, the eccentricity distribution, and the fraction of multiple planetary systems are but a few of the many outstanding questions to be addressed in this young field.

We also encourage high-contrast imaging campaigns to include more A-type stars in their surveys.  Five planets have already been directly imaged orbiting A stars: HR 8799 b, c, and d (\citealt{Marois:2008p18841}), Formalhaut b (\citealt{Kalas:2008p18842}), and $\beta$ Pic b (\citealt{Lagrange:2009p14794}).  Moreover, \citet{Marois:2008p18841} made their discovery after observing only a few early-type stars, in contrast to the many hundreds of late-type stars that have yielded null detections.   These results, combined with the higher inner planet occurrence rate we measure, suggest that more planet-hunting imaging surveys in this mass regime will yield fruitful results.

 \acknowledgments
We extend our gratitude to the many CAT observers who have helped with this project over the years, including Chris McCarthy, Raj Sareen, Howard Isaacson, Joshua Goldston, Bernie Walp, Julia Kregenow, Jason Wright, and Shannon Patel.  We also gratefully acknowledge the efforts and dedication of the Lick Observatory staff, and the time assignment committee of the University of California for their generous allocations of observing time.  B. P. B. thanks Chris Beaumont and Eric Ford for productive discussions about MCMC techniques.  J. A. J. is an NSF Astronomy and Astrophysics Postdoctoral Fellow and acknowledges support from the NSF grant AST-0757887.  We appreciate funding from NASA grant NNG05GK92G (to G. W. M.).  G. W. H. acknowledges long-term support from NASA, NSF, Tennessee State University, and the state of Tennessee through its Centers of Excellence program.  D. A. F. is a Cottrell Science Scholar of Research Corporation and acknowledges support from NASA grant NNG05G164G that made this work possible.  This research made use of the SIMBAD database operated at CSD, Strasbourg France, and the NASA ADS database.

%{\it Facility:} \facility{Shane ()}, 

\newpage

% \bibliographystyle{apj}
% \bibliography{bibfile_v3}

\end{document}